\documentclass[pre,amsmath,amssymb,reprint,superscriptaddress]{revtex4-2}

\usepackage[dvipdfmx]{graphicx}
\usepackage{epsf}
\usepackage{bm}
\usepackage{physics}
\usepackage{lipsum}
\usepackage[normalem]{ulem}
\usepackage{txfonts}
\usepackage{color}
\def\gs{\gtrsim}

\newcommand*{\etal}{\textit{et al}. }

\begin{document}
\title{Unraveling the glass-like dynamic heterogeneity in ring polymer melts:\\ From semi-flexible to stiff chain}

\author{Shota Goto}
\affiliation{Division of Chemical Engineering, Department of Materials Engineering Science, Graduate School of Engineering Science, Osaka University, Toyonaka, Osaka 560-8531, Japan}

\author{Kang Kim}
\email{kk@cheng.es.osaka-u.ac.jp}
\affiliation{Division of Chemical Engineering, Department of Materials Engineering Science, Graduate School of Engineering Science, Osaka University, Toyonaka, Osaka 560-8531, Japan}

\author{Nobuyuki Matubayasi}
\email{nobuyuki@cheng.es.osaka-u.ac.jp}
\affiliation{Division of Chemical Engineering, Department of Materials Engineering Science, Graduate School of Engineering Science, Osaka University, Toyonaka, Osaka 560-8531, Japan}

%%%%%%%%%%%%%%%%%%%%%%%%%%%%%%%%%%%%%%%%%%%%%%%%%%%%%%%%%%%%%%%%%%%%%%%%%%%%%%%%%%%%%%%%%%%%%%%%%%%%%%%%%%%%%%%%%%%%%
\date{\today}

%%%%%%%%%%%%%%%%%%%%%%%%%%%%%%%%%%%%%%%%%%%%%%%%%%%%%%%%%%%%%%%%%%%%%%%%%%%%%%%%%%%%%%%%%%%%%%%%%%%%%%%%%%%%%%%%%%%%%
\begin{abstract}
Ring polymers are an intriguing class of polymers with unique physical
 properties, and understanding their behavior is important for
 developing accurate theoretical models.
In this study, 
we investigate the effect of chain stiffness and monomer density on 
static and dynamic behaviors of ring polymer melts using molecular dynamics simulations. 
Our first focus is on the non-Gaussian parameter of center of
 mass displacement as a measure of dynamic heterogeneity, which is
 commonly observed in glass-forming liquids.
We find that the non-Gaussianity in the displacement
 distribution increases with the monomer density and stiffness of the
 polymer chains, suggesting that excluded volume interactions between centers of mass
 have a stronger effect on the dynamics of ring polymers. 
We then
 analyze the relationship between the radius of gyration and monomer
 density for semi-flexible and stiff ring polymers. 
Our results indicate that
 the relationship between the two varies with chain stiffness, which can
 be attributed to the competition between 
repulsive forces inside the ring and from adjacent rings.
Finally, we study the dynamics of bond-breakage virtually connected
 between the centers of mass of rings to analyze the exchanges of
 inter-molecular networks of bonds.
Our
 results demonstrate that the dynamic heterogeneity of bond-breakage is
 coupled with the non-Gaussianity in ring polymer melts, highlighting
 the importance of bond-breaking method in determining the
 inter-molecular dynamics of ring polymer melts.
Overall, our study provides insights into the fundamental mechanism
 of ring polymers and sheds light on the factors that govern their
 dynamic behavior.
\end{abstract}
%%%%%%%%%%%%%%%%%%%%%%%%%%%%%%%%%%%%%%%%%%%%%%%%%%%%%%%%%%%%%%%%%%%%%%%%%%%%%%%%%%%%%%%%%%%%%%%%%%%%%%%%%%%%%%%%%%%%%
\maketitle

\section{Introduction}

The dynamic properties of polymers melts are governed by structural features, 
such as the chain length $N$ and ``topological constraints''
(TCs)~\cite{gennes1979Scaling, doi1986Theory}.
In linear polymer melts, entanglement effects are common TCs and play a
key role in describing the $N$ dependence of the diffusion constant $D$.
However, defining and characterizing TCs in ring polymers 
is still challenging due to the absence of chain
ends~\cite{cates1986Conjectures, grosberg1993Crumpled, sakaue2011Ring,
halverson2014Melta, ge2016SelfSimilar, kim2021Intrinsic}.

In ring polymer melts, the simple picture of TCs is that 
they inhibit each other's dynamics due to
inter-ring ``threadings''~\cite{muller1996Topological, smrek2016Minimal,
landuzzi2020Persistence, stano2022Thread}.
As $N$ increases, the number of threading configurations also increases,
making it more difficult for the system to find the equilibrium configuration
to relax the threading.
The threading event of large $N$ rings suggests a slowing-down of the
dynamics, similar to the slow dynamics in glass-forming
liquids, where cage effects are imposed by 
the local density environment~\cite{donth2001Glass}. 
The concept of ``topological glass'' has been used to understand the
dynamics of ring polymer melts, highlighting the unique role of TCs
in these systems compared to the entanglements in linear
polymers~\cite{lo2013Topological, lee2015Slowing,
michieletto2016Topologically, michieletto2017Glassiness,
michieletto2017Ring, sakaue2018Topological, gomez2020Packinga,
smrek2020Active, chubak2020Emergence, michieletto2021Dynamical, chubak2022Active}.
Interestingly, 
techniques such as random pinning~\cite{michieletto2016Topologically, michieletto2017Glassiness} and
activeness~\cite{smrek2020Active, chubak2020Emergence, chubak2022Active}
have been introduced to enhance the glassiness in ring polymers through
molecular dynamics (MD) simulations.

Dynamic heterogeneity (DH) is a key concept used to describe the
significant slowing-down of glass-former liquids as they approach the glass
transition temperature~\cite{bohmer1996Dynamic, richert1998Dynamic, ediger2000Spatially}.
The slowing-down is accompanied by the collective structural relaxation
of spatially heterogeneous regions that exceeds the molecular
size~\cite{hurley1995Kinetic,
kob1997Dynamical, yamamoto1997Kinetic, donati1998Stringlike}.
DH is conventionally measured by the non-Gaussian parameter (NGP), i.e., the degree of the deviation
from the Gaussian distribution for the
molecular displacement within a given time
interval~\cite{kob1997Dynamical, hurley1996Non, shell2005Dynamic,
flenner2005Relaxation, saltzman2006NonGaussian, chaudhuri2007Universal}.
The NGP was utilized to quantify the non-Gaussianity in supercooled linear polymer
melts~\cite{aichele2003Polymerspecific, peter2009MD, pan2018Diffusion}.
In addition, 
we conducted calculations on the NGP for
linear polymer melts by MD simulations using the
Kremer--Grest (KG) bead-spring model~\cite{goto2021Effects}.
The chain lengths varied from $N=5$ to 
400, and the monomer density was set at $\rho=0.85$ (in the unit of $\sigma^{-3}$ using the size of the bead
$\sigma$).
Our findings revealed that a notable increase in the peak of the NGP as
$N$ increases.
This suggests 
that the dynamics of the system becomes spatially heterogeneous.
However, note that 
the mechanism of non-Gaussianity in linear polymer melts is due to
the enhanced mobility of chain ends, which is different
from the cage effects observed in glass-forming liquids.

Michieletto \etal conducted MD simulations of ring polymers using the
KG model and analyzed the center-of-mass (COM)
displacement distribution~\cite{michieletto2017Glassiness}.
They found that the non-Gaussian behavior was pronounced
even in the absence of random pinning fields, when the monomer
density $\rho$ increased with the chain length $N=500$.
This finding is consistent with
the experimental observation of polyethylene oxide ring melts by Br\'{a}s \etal\cite{bras2014Compact}
However, our previous study, which also used the same model for MD
simulations of ring polymer melts, showed that the NGP remained
quite small at all time regimes, even when the chain length was
increased up to $N=400$~\cite{goto2021Effects}.
It should be noted that the chain stiffness differed between the
two studies. 
Specifically, the bending
potential $\varepsilon_\theta(1+\cos\theta)$ (in the unit of energy
scale in the Lennard-Jones potential) acts on the bending angle
$\theta$ formed by three consecutive monomer beads
along the polymer chain (refer to Eq.~\eqref{eq:bending}).
Michieletto \etal utilized a stiff
ring chain with the bending energy of $\varepsilon_\theta = 5$ for
densities up to $\rho = 0.4$.
More recently, the glass-like slow dynamics has also been
demonstrated at low densities by increasing the chain stiffness up to 
$\varepsilon_\theta =20$~\cite{roy2022Effect}.
By contrast, we simulated
semi-flexible ring chains with $\varepsilon_\theta= 1.5$ at a
higher density of $\rho = 0.85$, which is the same as that used
in the MD study by Halverson \etal\cite{halverson2011Molecular,
halverson2011Moleculara, halverson2012Rheologya}

Thus, there is still much to be explored regarding the influence of
chain stiffness on DH in ring polymer melts.
To address this gap, 
we performed MD simulations using the KG model by
varying $\varepsilon_\theta$ and $\rho$.
Our analysis began by examining the NGP, and characterized the effect
of chain stiffness on the DH in ring polymer melts.
We also 
investigated the conformation of ring chains by analyzing the radius of
gyration, as well as asphericity and prolateness based on the diagonalization of the 
gyration tensor.
Additionally, we introduced the concept of inter-molecular bonds
virtually connected by ring
COM positions, which enabled us to investigate the rearrangement of
inter-molecular connectivity of ring polymers.
By combining the results obtained from these analyses, we aim to 
identify similarities and differences in the effects of chain
stiffness and monomer density on ring polymer dynamics.

\section{Model and Methodology}
% -- Models -- %

We employed MD simulations for ring polymer melts utilizing the KG model~\cite{kremer1990Dynamics}.
Each ring polymer was represented by $N$ monomer beads of mass $m$ and diameter $\sigma$.
Our system consisted of $M$ polymer chains contained within a three-dimensional cubic box with volume of $V$
and periodic boundary conditions.
All monomer beads were subject to 
three types of inter-particle potentials, namely:
the Lennard-Jones (LJ) potential, which acted between
all pairs of monomer beads,
\begin{equation}
  U_{\mathrm{LJ}}(\bm{r}) = 4 \varepsilon_{\mathrm{LJ}}
  \qty[ \qty(\frac{\sigma}{r})^{12} - \qty(\frac{\sigma}{r})^6 ]
  + C.
  \label{eq:LJ}
\end{equation}
Here $r$ and $\varepsilon_\mathrm{LJ}$ 
represent the distance between two monomer beads and the energy scale of LJ potential, respectively.
The LJ potential was truncated at the cut-off distance of $r_\mathrm{c} = 2^{1/6}\ \sigma$,
and the constant $C$ ensured that the potential energy shifted to zero at $r=r_\mathrm{c}$.
Additionally, two adjacent monomer beads along the chain also interacted
via the bond potential
\begin{equation}
  U_{\mathrm{bond}}(r) = -\frac{1}{2} K R_0^2
  \ln \qty[1 - \qty(\frac{r}{R_0})^2]
  \label{eq:FENE},
\end{equation}
for $r < R_0$, where $K$ and $R_0$ represent the spring constant and the 
maximum length of the bond, respectively.
Note that Eqs~\eqref{eq:LJ} and \eqref{eq:FENE} define
the finitely extensible nonlinear elastic (FENE) bond potential of the
KG model.
We adopted the values of $K=30\ \varepsilon_\mathrm{LJ}/\sigma^2$ and $R_0 =1.5\ \sigma$.
Lastly, 
we controlled 
the chain stiffness by incorporating a bending potential
\begin{equation}
  U_{\mathrm{bend}}(\theta) = \varepsilon_\theta 
  \qty[1 - \cos (\theta - \theta_0)],
  \label{eq:bending}
\end{equation}
where the bending angle $\theta$ is formed by three consecutive monomer beads
along the polymer chain.
In this study, 
we explored two bending energy cases: 
a semi-flexible chain with $\varepsilon_\theta = 1.5\varepsilon_{\mathrm{LJ}}$ and
a stiff chain with $\varepsilon_\theta = 5\varepsilon_{\mathrm{LJ}}$ both
with 
an equilibrium angle of $\theta_0 = 180^\circ$.

% -- Simulations -- %
We conducted MD simulations using the Large-scale Atomic/Molecular
Massively Parallel Simulator (LAMMPS)~\cite{plimpton1995Fast}.
Hereafter, the length, energy and time are conventionally represented in units of $\sigma$, $\varepsilon_\theta$ and 
$(m / \varepsilon_\mathrm{LJ})^{1/2}$, respectively.
Moreover, the temperature is also presented in units of $\varepsilon_\mathrm{LJ} / k_\mathrm{B}$, 
where $k_\mathrm{B}$ is Boltzmann constant.

We fixed the temperature $T$, chain length $N$, number of chains $M$ as $T=1.0$ and
$N=400$, and $M=100$, respectively.
During all simulations,
the temperature was controlled using 
the Nos\'{e}--Hoover thermostat, with a time step of $\Delta t = 0.01$.
We varied the monomer density $\rho\sigma^3$ ($= NM\sigma^3/V$) as $0.1$, $0.3$, $0.4$,
$0.5$, and $0.55$ both for the semi-flexible and stiff chains.
Besides, we adopted the monomer density $\rho = 0.85$ for the semi-flexible
chain with $\varepsilon_\theta = 1.5$, which was a common choice for MD simulations 
both for linear~\cite{hsu2016Static, hsu2017Detailed} and 
ring~\cite{halverson2011Molecular,
halverson2011Moleculara, halverson2012Rheologya, parisi2021Nonlinear, goto2021Effects} polymers.
It should be noted that 
a stiff chain system with $\varepsilon_\theta = 5$ displayed nematic
ordering when the monomer densities exceeded $\rho=0.55$, which is in
agreement with the recent MD simulations reported in ref.~\citenum{tu2023Unexpected}.
Therefore, the system of $\varepsilon_\theta = 5$ at $\rho=0.85$ was excluded from the analysis.
For each combination of $\varepsilon_\theta$ and
$\rho$ with the chain length $N=1,000$, 
we calculated the Kuhn length $l_\mathrm{k}$ 
using
$l_\mathrm{k} = \langle R^2 \rangle/l_\mathrm{b} (N-1)$,
in the linear chain melt~\cite{faller1999Local}.
Here, $\langle R^2 \rangle$ represents the mean square
end-to-end distance of the chain, and $l_\mathrm{b} \simeq 0.97$ denotes
the average distance between two
neighboring beads 
in the KG model.
Another important characteristic is the entanglement length
$N_\mathrm{e}$, which we determined through the primitive path
analysis~\cite{sukumaran2005Identifying, hagita2021Effect}.
The values of $l_\mathrm{k}$ and $N_\mathrm{e}$ are presented in Table~\ref{tbl:lk_Ne}.
Note that in previous studies, $l_\mathrm{k}$ was reported to be 
$l_\mathrm{k}\simeq 2.79$
for $\varepsilon_\theta=1.5$ at
$\rho=0.85$ and 
$l_\mathrm{k}\simeq 10$
for $\varepsilon_\theta=5$ at $\rho=0.1$,
respectively~\cite{halverson2011Molecular, michieletto2017Glassiness}.
Additionally, $N_\mathrm{e}$ was reported to be 
$N_\mathrm{e} \simeq 28$ for
$\varepsilon_\theta=1.5$ at $\rho=0.85$ and 
$N_\mathrm{e}
\simeq 40$ for $\varepsilon_\theta=5$ at
$\rho=0.1$, respectively~\cite{everaers2004Rheology, michieletto2016Topologically}.
However, 
we encountered difficulties in estimating $N_\mathrm{e}$ at the density
$\rho = 0.1$ both for both $\varepsilon_\theta = 1.5$ and 5 due to the
absence of entanglement effects with $N = 1,000$.

\begin{table}
  \caption{Kuhn length $l_\mathrm{k}$ and entanglement length
 $N_\mathrm{e}$ by varying bending energy $\varepsilon_\theta$ and
 monomer dnesity $\rho$. (*: No entanglement effects were observed.)}
  \label{tbl:lk_Ne}
  \begin{tabular}{cccc}
    \hline
    $\varepsilon_\theta$  & $\rho$ & $l_\mathrm{k}$ & $N_\mathrm{e}$ \\
    \hline
    1.5 & 0.1 & 4.1 & *  \\
    1.5 & 0.3 & 3.8 & 121  \\
    1.5 & 0.4 & 3.7 & 85  \\
    1.5 & 0.5 & 3.0 & 60  \\
    1.5 & 0.55 & 3.0 & 59  \\
    1.5 & 0.85 & 2.8 & 28  \\
    5 & 0.1 & 10 & *  \\
    5 & 0.3 & 7.8 & 32  \\
    5 & 0.4 & 7.2 & 24  \\
    5 & 0.5 & 6.4 & 19  \\
    5 & 0.55 & 6.4 & 15  \\
    \hline
  \end{tabular}
\end{table}

% -- MSD NGP -- %
\begin{figure*}[t]
\centering
\includegraphics[width=0.7\textwidth]{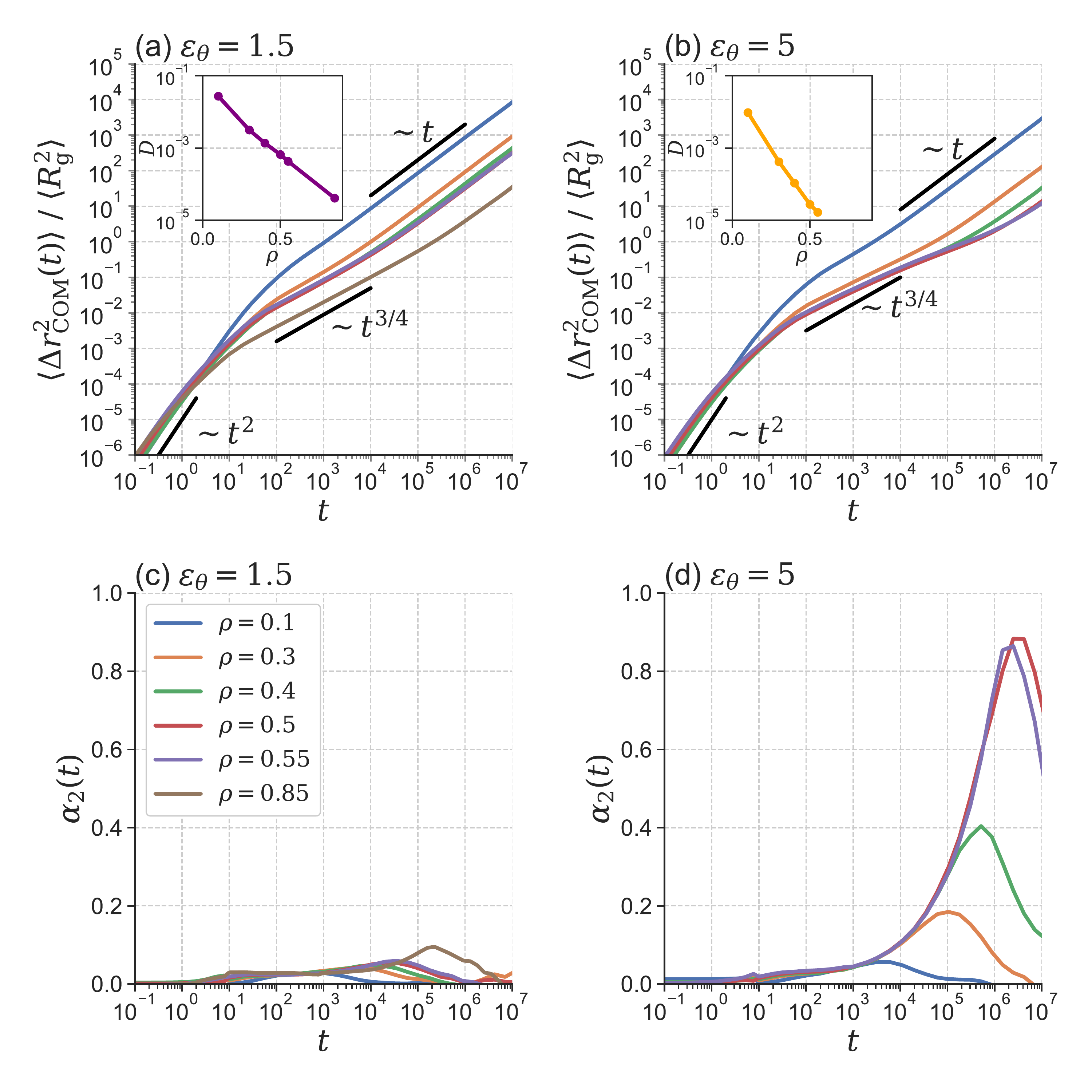}
\caption{Monomer density $\rho$ dependence of 
MSD $\langle \Delta r^2_\mathrm{COM}(t) \rangle$ and NGP $\alpha_2(t)$
 for $\varepsilon_\theta = 1.5$ [(a) and (c)] and for
 $\varepsilon_\theta = 5$ [(b) and (d)],
 respectively.
Note that MSD is scaled by mean square gyration of radius $\langle R_\mathrm{g}^2 \rangle$.
In (a) and (b), the ballistic, sub-diffusive, and diffusive behaviors,
 $\langle\Delta r^2_\mathrm{COM}(t)\rangle \sim t^\alpha$, are
 represented by black lines with $\alpha=2$, $3/4$, and $1$, respectively.
 Insets of (a) and (b): semi-log plots of the diffusion constant $D$ as a
 function of the monomer density $\rho$, respectively.
Note that the monomer density $\rho = 0.85$ was analyzed only for $\varepsilon_\theta = 1.5$.}
\label{fig:MSD-NGP}
\end{figure*}

\section{Results and Discussion}

\subsection{Mean Square Displacement and Non-Gaussian Parameter}

% -- asphericity -- %
\begin{figure*}[t]
  \centering
  \includegraphics[width=\textwidth]{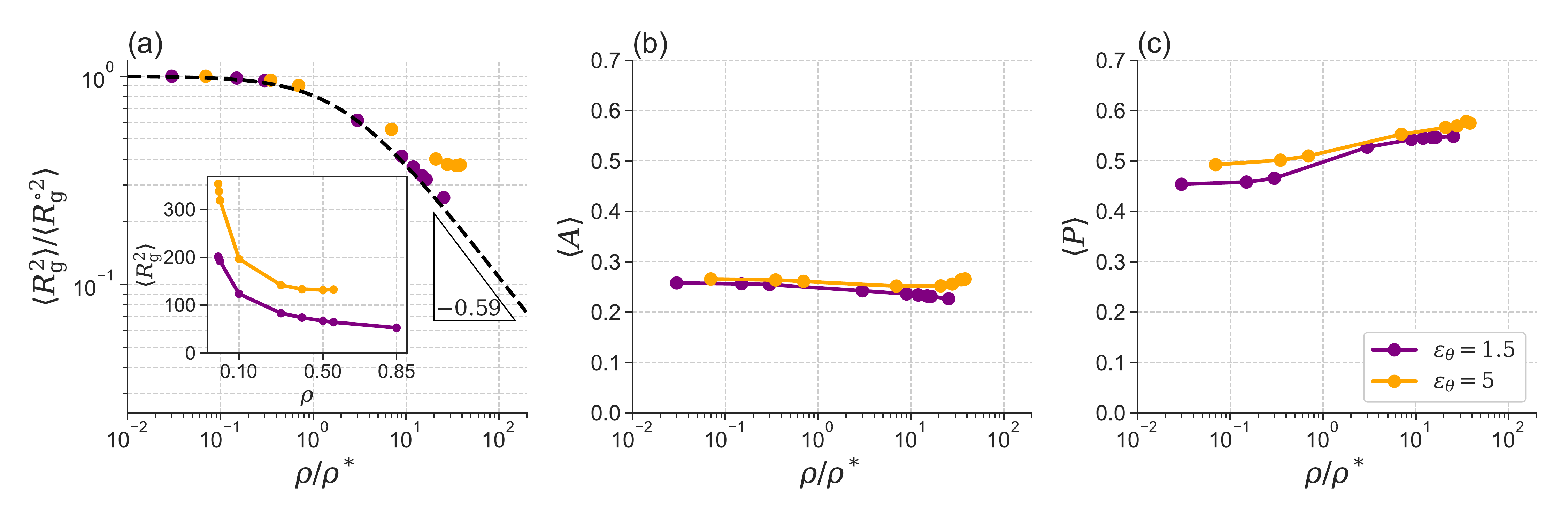}
  \caption{Monomer density $\rho$ dependence of chain conformation characteristics:
  (a) mean square radius of gyration $\langle R_\mathrm{g}^2 \rangle / \langle R_\mathrm{g}^{\circ 2} \rangle$, 
  (b) asphericity $A$, and 
  (c) prolateness $P$. 
In (a), the mean square radius of gyration is normalized by $\langle
 R_\mathrm{g}^{\circ 2} \rangle$, which represents the mean square
 radius of gyration at a density of $\rho=0.001$.
 The raw data of $\langle R_\mathrm{g}^2 \rangle$ as a
 function of $\rho$ is also shown in Inset of panel (a).
The black dotted line in (a) is the master curve, 
  $\langle R_\mathrm{g}^2\rangle / \langle R_\mathrm{g}^{\circ 2}\rangle = [1 + 0.45 (\rho /
 \rho^*)]^{-0.59}$.
In each panel, the density is scaled density $\rho^* = 3N/(4\pi \langle R_\mathrm{g}^{\circ 2}\rangle ^{3/2})$.
}
\label{fig:static_param}
\end{figure*}

We first analyzed the mean square displacement (MSD) of 
the COM of ring polymer chains and the
NGP of the COM displacement distribution.
The mean value of the even power of the COM displacement is defined by 
\begin{equation}
\langle \Delta r^{2n}_\mathrm{COM}(t) \rangle = \left\langle
						\frac{1}{M}\sum_{m=1}^M
						|\bm{R}_m(t)-
						\bm{R}_m(0)|^{2n}
						\right\rangle,\quad
(n=1, 2, \cdots),
\label{eq:MSD}
\end{equation}
where $\bm{R}_m(t)$ represents the COM position of $m$-th polymer chain
at time $t$.
Here, $\langle \cdots \rangle$ denotes an average over the initial time.
The second order with $n=1$ corresponds to the MSD.
Furthermore, the NGP for the center of mass (COM) displacement $\alpha_2(t)$ is
defined by
\begin{equation}
\alpha_2(t) = \frac{3}{5}\frac{\langle \Delta
 r_\mathrm{COM}^4(t)\rangle}{\langle \Delta
 r_\mathrm{COM}^2(t)\rangle^2} - 1.
\label{eq:NGP}
\end{equation}
The NGP is a typical quantity to characterize DH in glass-forming liquids, which 
measures the non-Gaussianity, i.e., the degree of the deviation of the distribution function of the 
COM displacement from the Gaussian form during the time interval $t$.

The results of MSD and NGP are
displayed in Fig.~\ref{fig:MSD-NGP} by changing the monomer density $\rho$ 
for $\varepsilon_\theta = 1.5$ [(a) and (c)] and $\varepsilon_\theta =
5$ [(b) and (d)], respectively.
As the monomer density $\rho$ increased, the diffusion of ring polymer chains
significantly slowed down both for $\varepsilon_\theta = 1.5$ and 5.
Moreover, at higher densities, the MSD exhibits a sub-diffusive behavior with 
$\langle \Delta r^{2}_\mathrm{COM}(t) \rangle \sim t^{3/4}$, followed by
diffusion behavior observed at displacements larger than mean
square gyration of radius $\langle R_\mathrm{g}^2\rangle$.
The COM diffusion constant $D$ was determined from the Einstein relation, 
$
D= \lim_{t\to\infty} \langle \Delta r^{2}_\mathrm{COM}(t) \rangle / 6t$.
The monomer density $\rho$ dependence of $D$ for $\varepsilon_\theta =
1.5$ and 5 is shown in the insets
of Fig.~\ref{fig:MSD-NGP}(a) and (b), respectively.
The reduction in diffusion was
more pronounced for the stiff chains with $\varepsilon_\theta = 5$ 
compared to semi-flexible chains with $\varepsilon_\theta = 1.5$ 
at time scales corresponding to the onset of the diffusion process at the same
monomer density.
These observations are consistent with the calculations by Michieletto
\etal\cite{michieletto2017Glassiness} and Halverson
\etal\cite{halverson2011Moleculara}
The mean square radius of gyration $\langle R_\mathrm{g}^2 \rangle$ will be 
discussed in the next subsection with respect to Fig.~\ref{fig:static_param}.

As demonstrated in Fig.~\ref{fig:MSD-NGP}(c),
the NGP's value of semi-flexible ring chains with $\varepsilon_\theta = 1.5$
remained relatively small ($\alpha_2(t) \lesssim 0.1$) at all investigated times and densities.
This suggests that the distribution of the COM displacement
$|\bm{R}_m(t)-\bm{R}_m(0)|$ follows a Gaussian distribution, which was
previously reported in
our work.~\cite{goto2021Effects}
The observation of Gaussian behavior in semi-flexible ring polymers, even at
the dense melt density of $\rho=0.85$, is noteworthy and
provides a unique perspective on the dynamics of ring polymers.
By contrast, 
for stiff ring chains, 
the increase in $\alpha_2(t)$ was more significant, showing peaks in a
long time regime that approximately corresponded to the onset time scale of the diffusive
behavior with $\langle \Delta r^{2}_\mathrm{COM}(t) \rangle \sim t$, as
demonstrated in Fig.~\ref{fig:MSD-NGP}(b) and (d).
Namely, the DH was found to be more pronounced in stiff ring chains 
with $\varepsilon_\theta = 5$, similar to common observations in glass-forming liquids.
An analogous glass-like heterogeneous dynamics was 
reported by Michieletto \textit{et al.}, who analyzed the displacement
distribution of stiff ring chains with $\varepsilon_\theta = 5$
up to $\rho=0.4$ with $N=500$~\cite{michieletto2017Glassiness}.
Therefore, the contracting observations in the NGP call for further
investigations into the 
COM mobility, which could entail significant differences between semi-flexible and stiff ring chains.

% -- rdf -- %
\begin{figure*}[t]
    \centering
    \includegraphics[width=0.7\textwidth]{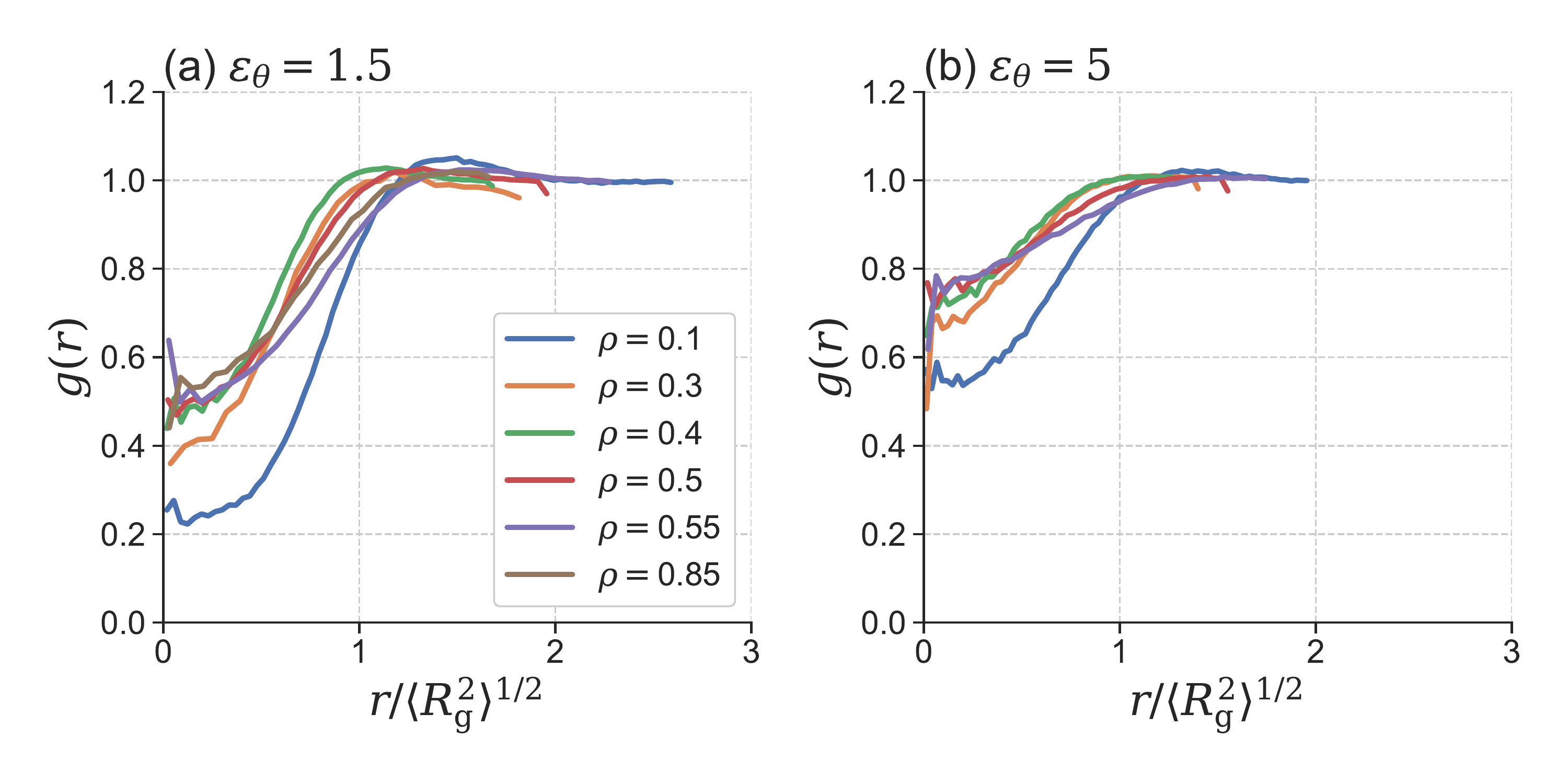}
    \caption{Radial distribution function $g(r)$ for COM of ring
 polymers as a function of the scaled distance $r/\langle R_\mathrm{g}^2\rangle^{1/2}$.
Results are shown for $\varepsilon_\theta = 1.5$ (a) and
 $\varepsilon_\theta = 5$ (b).
    }
    \label{fig:rdf}
\end{figure*}

\subsection{Conformation of the Ring Chains: Radius of Gyration, Asphericity and Prolateness}

% -- bond number -- %
\begin{figure*}[t]
    \centering
    \includegraphics[width=\textwidth]{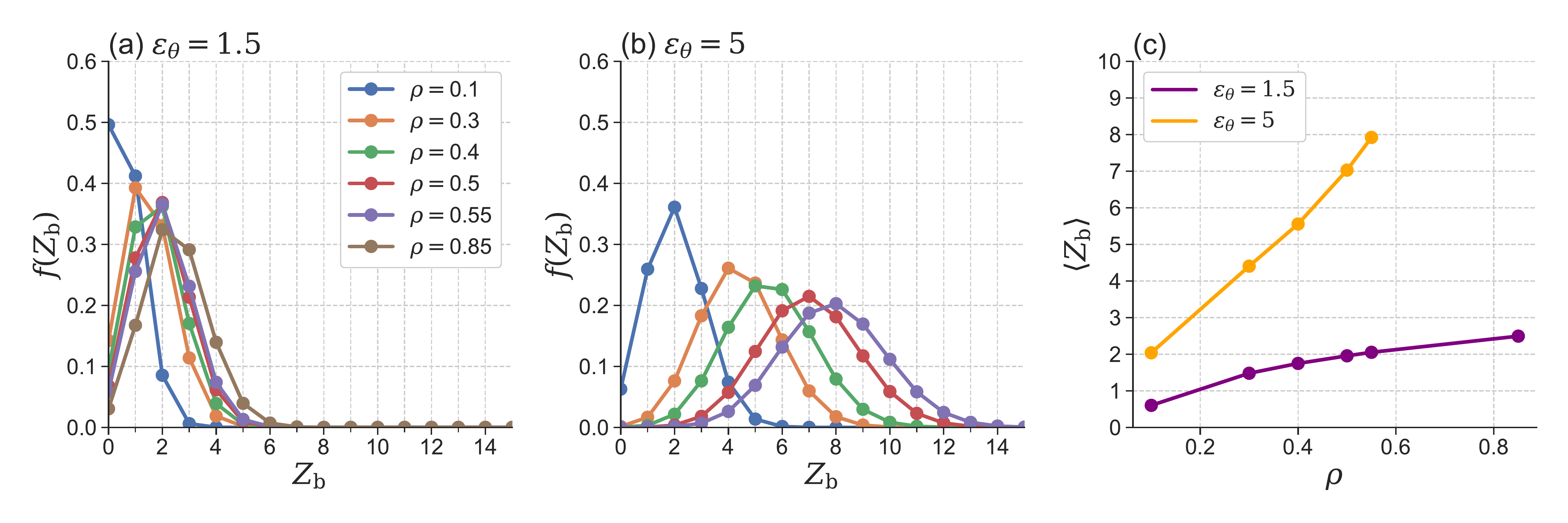}
    \caption{Probability distributions of the number of virtual bonds,
 $f(Z_\mathrm{b}$), 
    for ring polymers of $\varepsilon_\theta = 1.5$ (a) and 
 $\varepsilon_\theta = 5$ (b).
  The virtual bonds are defined based on Eq.~\eqref{eq:A1}.
Panel (c) shows the monomer density $\rho$ dependence of the mean value of $Z_\mathrm{b}$.
}
    \label{fig:pdf(N_b)}
\end{figure*}

% -- visualization --% 
\begin{figure}[t]
  \centering
  \includegraphics[width=0.5\textwidth]{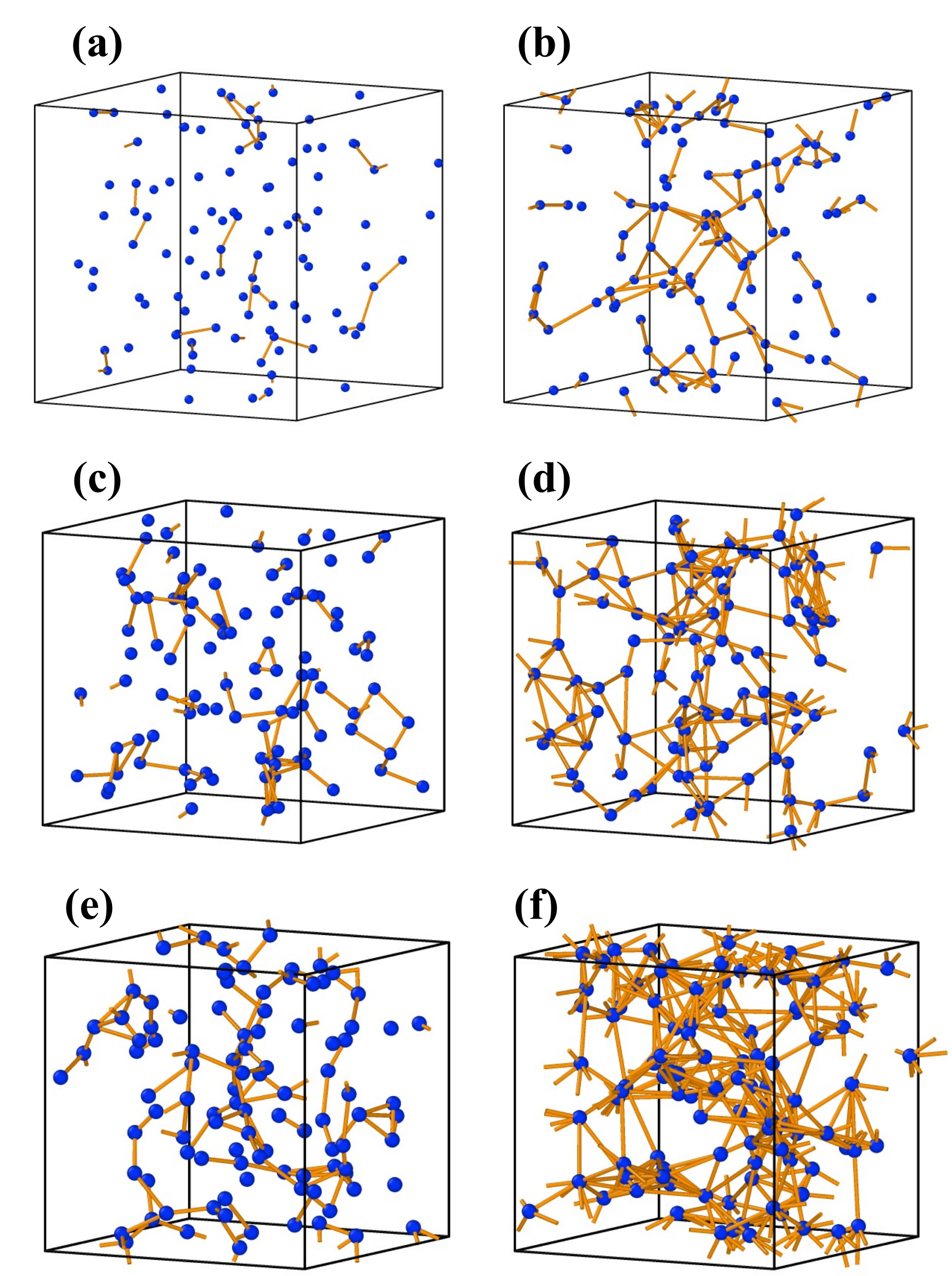}
  \caption{
  Visualization of virtual bonds (yellow lines) between the COM of rings (blue spheres)
  for $\varepsilon_\theta = 1.5$ [(a), (c) and (e)] and $\varepsilon_\theta = 5$ [(b), (d) and (f)].
  The monomer density $\rho$ increases as $\rho = 0.1$, 0.3, and 0.5 from top to bottom.
  }
  \label{fig:vis_ntw}
\end{figure}

% -- gyration tensor based analysis -- %
It is important to examine the details regarding the conformation of rings and its relationship with
the DH both for semi-flexible and stiff chain.
The radius of gyration provides a measure of the size of polymer chains.
To gain a more sophisticated understanding of the shapes, the principal components the gyration tensor $\bm{I}$
can be utilized, which allows for examination of 
the asphericity and prolateness of the polymer
chains~\cite{aronovitz1986Universal, rudnick1986Aspherity,
gaspari1987Shapes, jagodzinski1992Universal}.
The gyration tensor for each ring chain is defined as $I_{\alpha \beta}
= N^{-1} \sum_{i=1}^{N}\sum_{j=1}^{N} (\alpha_i - \alpha_j)(\beta_i -
\beta_j)$,
where 
$\alpha_i$ represents the $\alpha$ element of $i$-th bead with 
$\alpha$, $\beta$ $(=x, y, z)$.
Here, the square radius of gyration $R_\mathrm{g}^2$ can be calculated as the summation of 
the eigenvalues $\lambda_i$ $(i=1, 2, 3)$ of the gyration tensor
$\bm{I}$ as 
$
R_\mathrm{g}^2 = \lambda_1 + \lambda_2 + \lambda_3,
$
where 
the principle axes of inertia are chosen such that the diagonal elements are ordered 
as $\lambda_1 \ge \lambda_2 \ge \lambda_3$.
Furthermore, the asphericity $A$ and prolateness $P$ were calculated
from the following equations:
\begin{align}
  & A=\frac{\left(\lambda_1-\lambda_2\right)^2+\left(\lambda_1-\lambda_3\right)^2+\left(\lambda_2-\lambda_3\right)^2}{2\left(\lambda_1+\lambda_2+\lambda_3\right)^2}, \\
  & P=\frac{\left(2 \lambda_1-\lambda_2-\lambda_3\right)\left(2 \lambda_2-\lambda_1-\lambda_3\right)\left(2 \lambda_3-\lambda_1-\lambda_2\right)}{2\left(\lambda_1^2+\lambda_2^2+\lambda_3^2-\lambda_1 \lambda_2-\lambda_1 \lambda_3-\lambda_2 \lambda_3\right)^{3 / 2}}.
  \label{eq:AandP}
\end{align}
The asphericity takes on values of $0 \le A \le 1$, where $A=0$
corresponds to spherically symmetric object, and 
$A=1$ corresponds to a polymer that is fully extended to form a rigid
rod shape.
The prolateness $P$ is bounded between $-1$ and $1$, where $P=-1$
represents a fully oblate object such as a disk, 
and $P=1$ represents a prolate object in the shape of a rigid rod.
The gyration tensor was calculated for each chain and the time
evolutions of $R_\mathrm{g}^2$, $A$, $P$ were computed.
The mean values $\langle R_\mathrm{g}^2 \rangle,\ \langle A \rangle$ and 
$\langle P \rangle$ were evaluated 
by taking the average of these quantities over the time series data for each chain.
It should be noted that
the analysis of the gyration tensor was performed in various simulations
of ring polymers~\cite{bishop1985Shape, bishop1988Polymer, jagodzinski1992Universal,
zifferer2001Monte, alim2007Shapes, rawdon2008Effect,
reigh2013Concentration, cai2022Conformation}.

Reigh and Yoon reported a universal scaling behavior of $\langle R_\mathrm{g}^{2}\rangle \sim \rho^{-0.59}$ for long ring
polymers by Monte Carlo simulation of a lattice model~\cite{reigh2013Concentration}.
This exponent $-0.59$ is significantly different from the value of
$-0.25$ observed 
for linear polymers, which was a well-established prediction based on scaling arguments.
This observation suggests
ring chains form more compact conformations than linear chains.
More recently, Cai \etal performed 
MD simulations of ring polymers using the KG model by varying
chain lengths $N$ up to 5120, and reported the same scaling
behavior of $\langle R_\mathrm{g}^{2}\rangle \sim \rho^{-0.59}$~\cite{cai2022Conformation}.
The master curve was heuristically proposed and given by
\begin{equation}
  \langle R_\mathrm{g}^2\rangle / \langle R_\mathrm{g}^{\circ 2} \rangle= [1 + 0.45 (\rho / \rho^*)]^{-0.59},
  \label{eq:rg_mastercurve}
\end{equation}
where 
$\langle R_\mathrm{g}^{\circ 2}\rangle$ denotes the mean square
radius of gyration in the dilute solutions.
In addition, $\rho^*$ corresponds to 
the overlap density defined by $\rho^* = 3N/(4\pi \langle R_\mathrm{g}^{\circ 2}\rangle ^{3/2})$.
They also compared their simulation results with available experimental
data, and found good agreement between simulations and experiments.
Note that the ring polymer chains in their simulations were fully flexible
because they did not incorporate any bending potentials.

Figure~\ref{fig:static_param}(a) shows 
the relative mean square radii of gyration $\langle
R_\mathrm{g}^{2}\rangle/ \langle R_\mathrm{g}^{\circ 2}\rangle$ as a
function of the
scaled density $\rho/\rho^*$.
We estimated $\langle R_\mathrm{g}^{\circ 2}\rangle$ 
as the value of mean square radii of gyration $\langle
R_\mathrm{g}^{2}\rangle$  
both for semi-flexible and stiff chains at a density of $\rho = 0.001$.
This density corresponds to a sufficiently low scaled density ($\rho /
\rho^* < 10^{-1}$),
making it appropriate to consider $\rho = 0.001$ as a dilute solution
both for semi-flexible and stiff ring polymers.
The data for semi-flexible rings with $\varepsilon_\theta = 1.5$ follow
the master curve given by Eq.~\eqref{eq:rg_mastercurve}.
However, a deviation from Eq.~\eqref{eq:rg_mastercurve} was
observed for stiff rings with
$\varepsilon_\theta = 5$, indicating that $\langle
R_\mathrm{g}^{2}\rangle$
of stiff ring chains decreases slightly slower than that of semi-flexible
ring chains as the density is increased beyond $\rho/\rho^* \gs 10$.
The inset in Fig. \ref{fig:static_param} (a) shows the density $\rho$ dependence 
of the mean square radii of gyrations $\langle R_\mathrm{g}^2 \rangle$.
This represents that the stiff rings are larger than
semi-flexible ones in all densities $\rho$.

Figure~\ref{fig:static_param}(b) and (c) show the 
average asphericity $\langle A\rangle $ and average
prolateness $\langle P\rangle$, respectively, as functions of $\rho/\rho^*$.
Interestingly, we found that 
the values of $\langle A\rangle$ and 
$\langle P\rangle $ approached
saturation regardless of the bending energy $\varepsilon_\theta$.
In particular, the relatively small values of $\langle A\rangle \simeq 0.2$
suggest that the ring polymer adopt globular conformations, which remain
valid across the densities examined.
However, slightly large values of $\langle P \rangle \simeq 0.5$
indicate that the rings extend moderately in the direction of longest inertia axis.
These imply that 
the shape of the rings is alomost spherical and relatively insensitive to both 
$\varepsilon_\theta$ and $\rho$, provided that the chain length is
sufficiently long compared to the Kuhn length scale.

\subsection{Inter-penetration of Ring Chains}

% -- bond correlation -- %
\begin{figure*}[t]
  \centering
  \includegraphics[width=\textwidth]{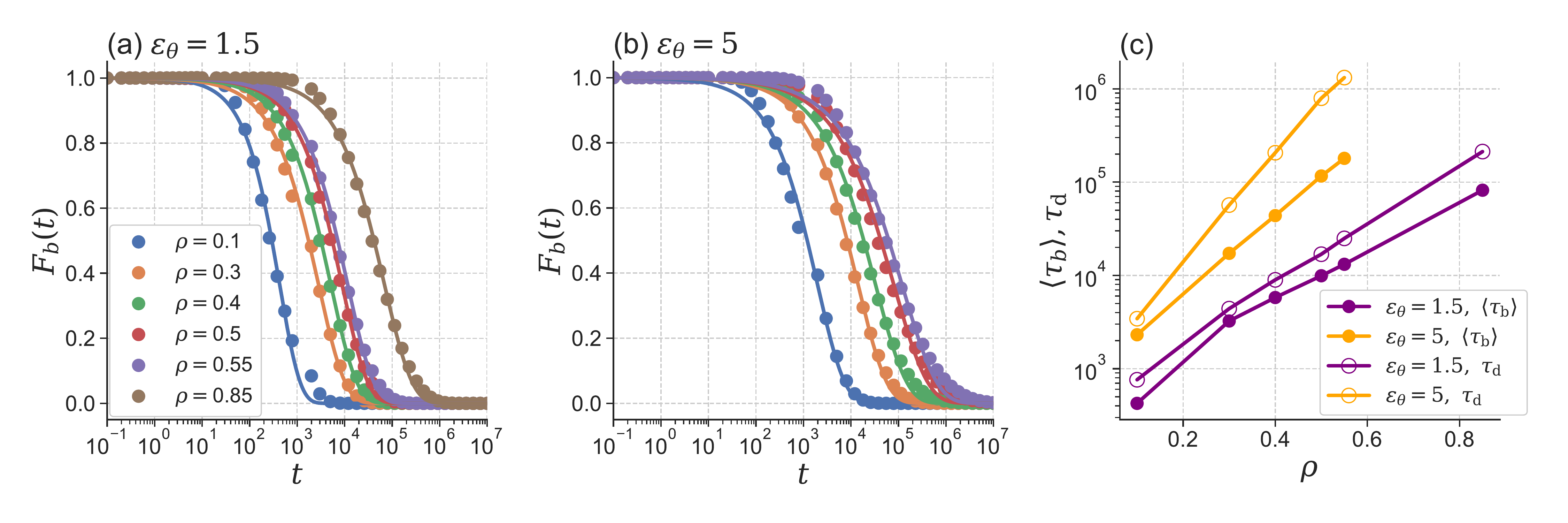}
  \caption{Monomer density dependence of 
the bond correlation function $F_\mathrm{b}(t)$
 for $\varepsilon_\theta = 1.5$ (a) and 
 $\varepsilon_\theta = 5$ (b), respectively.
The solid line represents the fitting result obtained using the stretched
 exponential function, $F_\mathrm{b}(t)\approx
 \exp[-(t/\tau_\mathrm{b})^{\beta}]$.
Panel (c) shows the monomer density $\rho$ dependence of the
 average relaxation time $\langle \tau_\mathrm{b}\rangle$ of the bond
 correlation function $F_\mathrm{b}(t)$ and the diffusion time 
 $\tau_\mathrm{d} = \langle R_\mathrm{g}^2\rangle / 6D$.
 }
  \label{fig:f_b}
\end{figure*}

As shown in Fig.~\ref{fig:static_param}, 
while the shape of the
polymer remained largely unchanged on average, there was a
increase in the mean square radius of
gyration $\langle R_\mathrm{g}^2 \rangle$ when the chain
stiffness increased to $\varepsilon_\theta = 5$.
This suggests that the inter-molecular connectivity of ring chains
may differs significantly between semi-flexible and stiff chains.
To explore this further, we calculated the radial distribution function for 
the COM of ring chains $g(r)$, and the results are presented in
Fig.~\ref{fig:rdf} as a function of the scaled length of $r/ \langle R_\mathrm{g}^2\rangle^{1/2}$.

As observed in Fig.~\ref{fig:rdf}, $g(r)$ allowed us to characterize the
degree of inter-penetration of ring polymers.
In fact, 
we did not observe a pronounced peak of 
$g(r)$, but instead found finite values at the length scale of $r <
\langle R_\mathrm{g}^2\rangle^{1/2}$, indicating that there is some degree of
inter-penetration between the ring chains.
The $g(r)$ became more broad with increasing the monomer density, suggesting
that the chains are less separated from each other.
Similar results of $g(r)$ were reported for flexible chains without the 
bending potential by Cai \etal\cite{cai2022Conformation}
Additionally, as shown in Fig.~\ref{fig:rdf}(b), the degree of the
inter-penetration became more significant as
the bending energy increased to $\varepsilon_\theta=5$.
This observation is consistent with the larger mean square radius of gyration
$\langle R_\mathrm{g}^2 \rangle$ of stiff rings with $\varepsilon_\theta=5$
compared to that of semi-flexible rings of 
$\varepsilon_\theta=1.5$ at the same density $\rho$.
The difference in $\langle R_\mathrm{g}^2 \rangle$ is also evident in Fig.~\ref{fig:static_param}, where the
curve for $\varepsilon_\theta=5$ is shifted to higher values of
$\rho/\rho^*$ compared to $\varepsilon_\theta=1.5$.
These results suggest that 
the competition between repulsive forces inside the ring and from adjacent rings
plays a crucial role in determining the loop structure.
While sufficiently semi-flexible polymers tend to be more compact
because the repulsion between neighboring rings overcomes the monomer bead
repulsion inside a single chain, 
the stiff polymers tend to expand due to the long Kuhn length,
leading to the inter-penetration of rings.

To analyze the number of inter-molecular connectivity,
we considered virtually connected bonds between the COM of ring chains.
In particular, for two polymers $i$ and $j$ with the COM positions $\bm{r}_i$ and $\bm{r}_j$,
they were considered to be virtually bonded if 
\begin{equation}
r_{ij} < A_1 \langle R_\mathrm{g}^2 \rangle^{1/2},
\label{eq:A1}
\end{equation}
with the value of $A_1=1$.
Here, $r_{ij} = |\bm{R}_i - \bm{R}_j|$ is the distance
between these COM.
For each polymer, 
the number of virtual bonds $Z_\mathrm{b}$, which represents
the static coordination number, was counted.
Figure~\ref{fig:pdf(N_b)} depicts the probability distribution $f(Z_\mathrm{b})$ 
for ring polymers of $\varepsilon_\theta = 1.5$ (a) and
$\varepsilon_\theta = 5$ (b) at varying the density $\rho$.
In the case of semi-flexible rings with $\varepsilon_\theta = 1.5$, the peak
was observed at around 2 for most densities, except for $\rho=0.1$, where
$Z_\mathrm{b}$ was predominantly 0, indicating that each ring chain was mostly
isolated and did not correlated with each other.
However, for stiff rings with $\varepsilon_\theta = 5$, we
observed an increase in the 
peak position and width of $f(Z_\mathrm{b})$ as the density $\rho$ increased.
The monomer density $\rho$ dependence of the mean value of
$Z_\mathrm{b}$ is shown in Fig.~\ref{fig:pdf(N_b)}(c).
Here, $\langle Z_\mathrm{b}\rangle$ can be evaluated by 
\begin{align}
\langle Z_\mathrm{b}\rangle = \int_0^{\langle R_\mathrm{g}^2
 \rangle^{1/2}} 4\pi r^2 \left(\frac{\rho}{N}\right) g(r) dr.
\end{align}
In cases of $g(r)=1$ and $\langle R_\mathrm{g}^2
 \rangle \sim \rho^{-0.6}$, 
$\langle Z_\mathrm{b}\rangle$ may exhibit a scaling behavior of $\langle Z_\mathrm{b}\rangle
\sim \rho \langle R_\mathrm{g}^2 \rangle^{3/2} 
\sim \rho^{0.1}$ at a fixed chain length $N$.
This suggests that $\langle Z_\mathrm{b}\rangle$ increases slowly as the density increases.
However, the presence of $g(r)<1$ for $r< \langle
 R_\mathrm{g}^2\rangle^{1/2}$, as observed in Fig.~\ref{fig:rdf} both for 
$\varepsilon_\theta=1.5$ and 5, leads to the deviates from the expected $\langle
 Z_\mathrm{b}\rangle \sim \rho^{0.1}$.
Notably, as shown in Fig.~\ref{fig:static_param}(a), $\langle R_\mathrm{g}^2 \rangle$ of $\varepsilon_\theta=5$ does
 not follow the $\langle R_\mathrm{g}^2 \rangle \sim \rho^{-0.6}$ scaling,
 resulting in a more pronounced increase in $\langle
 Z_\mathrm{b}\rangle$ with increasing the density.

Moreover, the spatial distribution of inter-molecular connectivity is visualized in Fig.~\ref{fig:vis_ntw}.
For semi-flexible ring polymers with $\varepsilon_\theta = 1.5$, bonds describing the 
connectivity of COM are sparse irrespective of the monomer density $\rho$.
In contrast, as the density increases,
ring polymers with higher stiffness
($\varepsilon_\theta = 5$) exhibit a stronger percolation, indicating 
a more interconnected network bond.
It is noteworthy that
there exists a critical coordination number around 3, beyond which the
linked ring polymers percolate through the entire system~\cite{michieletto2015Kinetoplast}.

\subsection{Rearrangements of Inter-molecular Connectivity}

% -- susceptibility -- %
\begin{figure*}[t]
  \centering
  \includegraphics[width=0.7\textwidth]{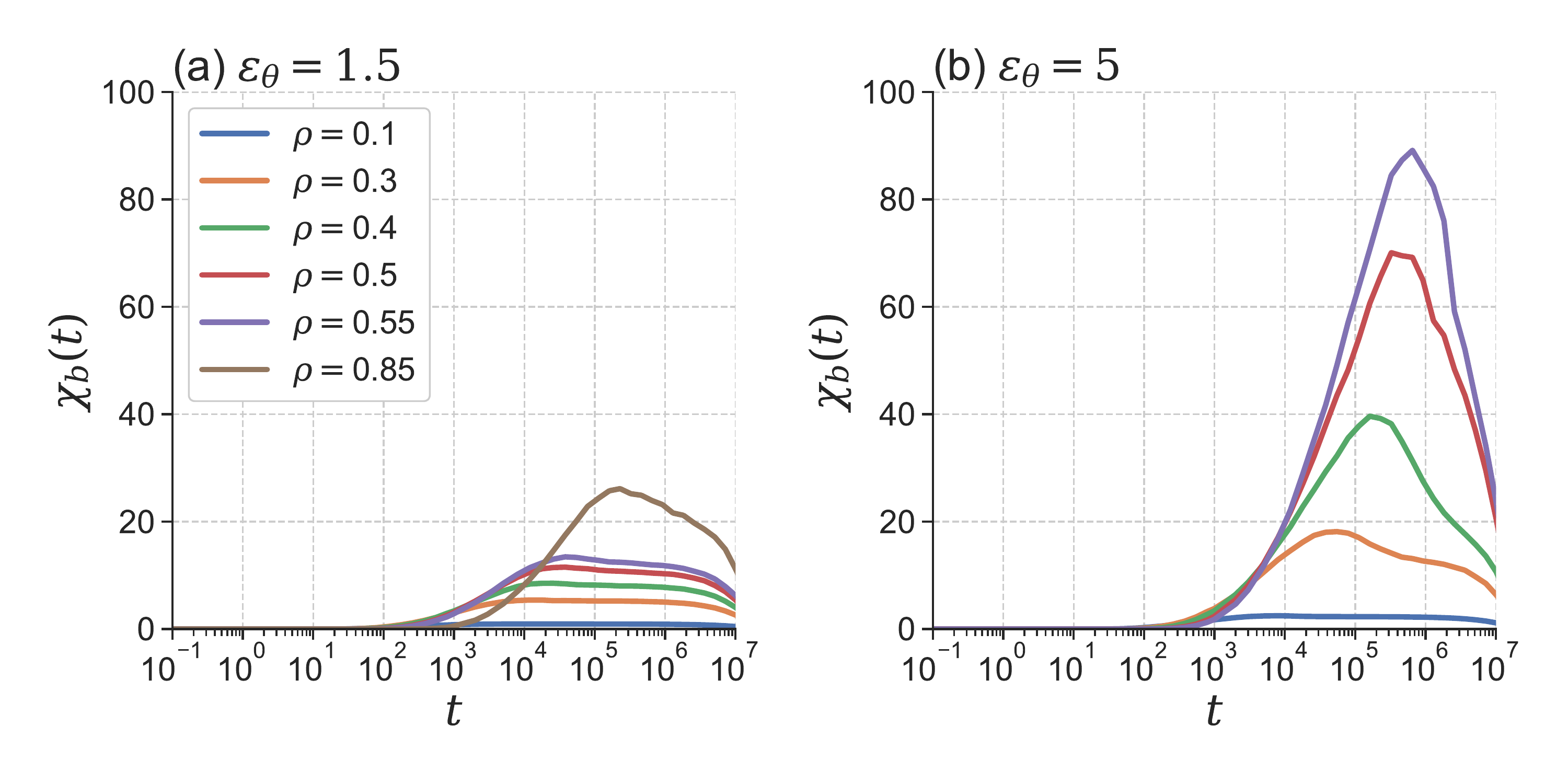}
  \caption{Monomer density dependence of 
the dynamic susceptibility of bond-breakage $\chi_\mathrm{b}(t)$ 
 for $\varepsilon_\theta = 1.5$ (a) and 
 $\varepsilon_\theta = 5$ (b),
 respectively.}
  \label{fig:chi_b}
\end{figure*}

% -- molecular bond -- %

To examine rearrangements of inter-molecular connectivity of ring
polymers, we analyzed the time evolution of virtual bonds.
This reflects the exchange of initially bonded neighbors 
because the COM motion breaks old bonds and forms new ones.
Although the average coordination number $\langle Z_\mathrm{b}\rangle$
may remain constant, the neighboring COMs will be replaced with
the new ones, thereby reshaping the cages around a tagged COM.
A similar methodology, known as the bond-breakage method, is used to
study the DH observed in
glass-forming liquids~\cite{yamamoto1997Kinetic, yamamoto1998Dynamics,
shiba2012Relationship, kawasaki2013Slow, shiba2016Unveiling}.

The virtual bond between two polymers $i$ and $j$ 
which had been counted to be formed at an initial time $0$
through Eq.~\eqref{eq:A1} was considered broken when
\begin{equation}
  r_{ij}(t) > A_2 \langle R_\mathrm{g}^2 \rangle^{1/2},
  \label{eq:A2}
\end{equation}
after a time interval of $t$.
To ensure bond-breaking insensitive to thermal fluctuations, 
the threshold value of $A_2=1.2$ was set slightly larger than $A_1$=1.
The total number of surviving bonds, $N_b(t)$, was calculated from the initial time $0$.
The bond correlation function,
$F_\mathrm{b}(t) =\langle N_b(t) /N_b(0) \rangle$, was obtained by averaging over 
the configurations at $t=0$.
Figure~\ref{fig:f_b} shows
the results of $F_\mathrm{b}(t)$ 
for $\varepsilon_\theta = 1.5$ (a) and $\varepsilon_\theta =5$ (b), respectively.
The characteristic time scale
of $F_\mathrm{b}(t)$ is related to that of the rearrangement of the
local coordination by the neighboring COMs, according to the definition of the bond.
The $F_\mathrm{b}(t)$ was fitted to the stretched exponential
function
$
F_b(t) = \exp\left[-\left(t/\tau_\mathrm{b}\right)^{\beta}\right],
$
where the exponent $\beta$ represents the degree of the deviation from
the exponential decay with $\beta =1$.
The average relaxation time $\langle \tau_\mathrm{b}\rangle $ was then
calculated from 
$
\langle \tau_\mathrm{b} \rangle = \int_0^\infty F_b(t) dt,
$
and estimated by $\langle
\tau_\mathrm{b}\rangle=(\tau_\mathrm{b}/\beta)\Gamma(1/\beta)$ with the
Gamma function $\Gamma(\cdots)$.
Figure~\ref{fig:f_b}(c) shows $\langle \tau_\mathrm{b}\rangle$ as a
function of the monomer density $\rho$.
Our results demonstrate the increase in the average relaxation time
$\langle \tau_\mathrm{b}\rangle$ of $F_\mathrm{b}(t)$ as the monomer density
$\rho$ increased, both for $\varepsilon_\theta = 1.5$ and $\varepsilon_\theta = 5$.
The increase in $\langle \tau_\mathrm{b}\rangle$ apparently obeys an
exponential trend as a function of $\rho$, except at
the dilute density of $\rho=0.1$ for $\varepsilon_\theta=1.5$, where the
average coordination number $\langle Z_\mathrm{b}\rangle$ is less than
1, indicating that polymer rings are nearly isolated (see Fig.~\ref{fig:pdf(N_b)}(c)).
Furthermore, we observed a more pronounced increase in $\langle
\tau_\mathrm{b}\rangle$ for stiff ring polymers with
$\varepsilon_\theta=5$, which is in accordance with the monomer density
$\rho$ dependence of $\langle Z_\mathrm{b}\rangle$ (see again Fig.~\ref{fig:pdf(N_b)}(c)).

Another significant time scale to consider is the diffusion time,
$\tau_\mathrm{d}$, defined as
$\tau_\mathrm{d} = \langle R_\mathrm{g}^2\rangle / 6D$, which 
corresponds to the time at which
the MSD reaches the length scale of the mean square radius of
gyration $\langle R_\mathrm{g}^2 \rangle$.
The monomer density dependence of $\tau_\mathrm{d}$ is illustrated in Fig.~\ref{fig:f_b}(c).
It is observed that 
for semi-flexible rings with $\varepsilon_\theta = 1.5$,
$\tau_\mathrm{d}$ increases in a similar manner to
$\langle\tau_\mathrm{b}\rangle$, while
for stiff rings with $\varepsilon_\theta = 5$, $\tau_\mathrm{d}$
exhibits a 
significant increase and 
becomes decoupled from $\langle\tau_\mathrm{b}\rangle$ as the density $\rho$ increases.
These observations suggest the COM diffusion of stiff rings is not
solely driven by 
local bond rearrangements, but requires a cooperative mechanism.

We then examined the collective effects of bond rearrangements
in ring polymers.
For this purpose, 
the dynamic susceptibility of bond-breakage was calculated by the fluctuation 
function of the number of broken-bonds at different time intervals, $t$~\cite{shiba2012Relationship}.
The number of the breakage-bond $B_i(t)$
between two times $0$ and $t$ for the $i$-th polymer was counted based on
the conditions given in Eqs.~\eqref{eq:A1} and \eqref{eq:A2}.
The degree of bond-breakage correlations can be characterized by the susceptibility $\chi_\mathrm{b}(t)$,
which is defined as 
\begin{equation}
  \chi_\mathrm{b}(t) = \frac{1}{M} \left \langle \sum_{i=1}^M \sum_{j=1}^M
    \delta B_i(t) \delta B_j(t) \right \rangle
    \label{eq:chi_b},
\end{equation}
where $\delta B_i(t) = B_i(t)/2 - \langle B(t) \rangle$ represents 
the deviation from the average number of broken bonds.
The average number of broken bonds can be 
estimated as $\langle B(t) \rangle = \langle \sum_{i=1}^M
B_i(t)/2\rangle / M$.
Note that the factor 1/2 avoided double-counting of the bond-breakage
between polymers $i$ and $j$.
Figure~\ref{fig:chi_b} illustrates the susceptibility of bond-breakage, $\chi_\mathrm{b}(t)$,
for different values of $\varepsilon_\theta$ and $\rho$.
For semi-flexible rings with $\varepsilon_\theta = 1.5$, the $\chi_\mathrm{b}(t)$ shows
relatively small values, whereas the peak of $\chi_\mathrm{b}$ became pronounced 
($\chi_\mathrm{b}\sim 30$) at the highest density $\rho = 0.85$ investigated.
In contrast, for 
stiff rings with $\varepsilon_\theta = 5$, 
the peaks show significant development with increasing monomer density,
particularly at the time regimes where the 
MSD nearly reaches the diffusive behavior. 
At the density of $\rho=0.55$, the peak height reaches $\chi_\mathrm{b}\sim 90$.
Therefore, the observed NGP behavior in Fig.~\ref{fig:MSD-NGP}(c) and (d)
is related to DH, which is also characterized by the bond-breakage
susceptibility, $\chi_\mathrm{b}(t)$.
Interestingly, 
the results of $\alpha_2(t)$ and $\chi_\mathrm{b}(t)$ suggest that 
ring polymers with $\varepsilon_\theta = 1.5$ exhibit spatial
homogeneous dynamics, even in the sub-diffusion regime.
In this perspective, the inter-chain interactions in semi-flexible
ring polymer melts display notable characteristics, while stiff ring
polymer melts exhibit interactions reminiscent of ``entanglements'' in linear
polymer melts.

Mei \etal have recently developed 
the polymer interaction site model (PRISM) 
as a microscopic theory for dense ring polymer melts~\cite{dell2018Intermolecular, mei2020Microscopic}.
This theory
proposes a partially inter-penetrating, two-step fractal structure model
for each ring chain and provides a master curve for the chain length $N$
dependence of the COM diffusion constant $D$.
Although the PRISM theory has shown good agreement with MD simulations data for 
semi-flexible ring polymers with
$\varepsilon_\theta=1.5$~\cite{halverson2011Moleculara}, deviations from
the master curve have been observed for stiff
rings of $\varepsilon_\theta = 5$~\cite{michieletto2017Glassiness}.
To gain a deeper understanding of the underlying mechanism of emergence
of DH in ring polymers melts,
a combined effort between theory and simulation may be necessary.
In particular, our MD simulation results analyzing DH can provide 
insights into the deviation from the master curves reported in
ref.~\citenum{mei2020Microscopic}, and may facilitate 
a generalization of the theory by incorporating an activated
hopping process~\cite{mei2021Theorya}.

\section{Conclusion}

In conclusion, our MD simulations of ring polymer melts using the
KG model have provided insights into the dynamics 
of semi-flexible and stiff ring chains. 
By analyzing the
NGP in the distribution of the COM displacement, we
have found that more stiff ring chains exhibit a peak in the NGP
in long time regimes, which increases with the monomer density. 
This suggests that the dynamics of stiff ring chains are affected by strong
inter-molecular interactions and that the motion of the COM is correlated
with each other. 
In contrast, more semi-flexible ring polymers
exhibit relatively small non-Gaussianity, indicating that the COM
mobility is almost uncorrelated with each other. 
The difference in
non-Gaussianity between the two types of ring polymers suggests that the
nature of the inter-molecular interactions changes significantly
depending on the degree of chain stiffness.

The behavior of the radius of gyration $R_g$ in relation to $\rho$
depends on the stiffness of the ring polymer chains. 
In the case of more semi-flexible rings, the $R_g$ follows a master curve described by
Eq.~\eqref{eq:rg_mastercurve}.
However, this curve does not apply to stiff ring polymer
melts. 
The deviation from the master curve can be explained by the
competition between the shrinkage caused by the excluded volume of
neighboring polymers and the expansion due to the chain stiffness. 
Specifically, semi-flexible ring polymers tend to adopt a compact
globule conformation due to the excluded volume interaction with their
neighbors, while more stiff rings expand due to the long Kuhn length.

We have also analyzed the dynamics of bond-breakage between
the COM of rings defined by using averaged radius of gyration, $\langle R_\mathrm{g}^2\rangle^{1/2}$. 
The network of virtual bonds in stiff rings are 
percolating,
while those in semi-flexible rings are sparsely
distributed. 
Furthermore, the results for the dynamic susceptibility of
bond-breakage are consistent with the non-Gaussianity in the
displacement distribution, indicating that the DH of
bond-breakage is coupled with the non-Gaussianity in diffusion in ring polymer melts.
Furthermore, it is crucial to investigate the dynamics of
ring-linear blend melts~\cite{jeong2017Relationa, borger2020Threading,
hagita2021Effect, oconnor2022Composite,
grest2023Entropic}.
In practical terms, the analysis of the bond-breakage is particularly well-suited for
this system, enabling the assessment of the inter-connectivity dynamics
of polymer chain COMs.

Threading is commonly discussed in ring polymer melts, but the
relationship with the bond-breakage dynamics remains
unclear. Further investigation into the properties of threading in ring
polymer melts with varying chain stiffness is warranted.
Finally, we have found that semi-flexible ring polymers exhibit
sub-diffusion yet Gaussian distribution with unique dynamics.
We suggest that the microscopic theory based on the PRISM for ring polymer melts will be 
useful for understanding the diffusion mechanisms of these systems.
Drawing on another crucial insight from ref.~\citenum{michieletto2015Kinetoplast},
we put forward the notion that the value of $\langle Z_\mathrm{b}\rangle
= 3$ acts as a threshold for the percolation of virtual bond
networks and the emergence of DH in ring polymers.
To gain deeper insights, further analysis is required, including the
cluster size distribution by varying the chain stiffness
$\varepsilon_\theta$ and extending the study to longer chain length $N$. 
Currently, we are pursuing the application of persistent homology
analysis to explore this perspective
further~\cite{landuzzi2020Persistence}.

\begin{acknowledgements}
The authors would like to thank Dr.~Takenobu Nakamura for valuable discussions.
This work was supported by 
JSPS KAKENHI Grant-in-Aid 
Grant Nos.~\mbox{JP22H04542} and ~\mbox{JP22K03550}.
This work was also partially supported by JST, the establishment of university
 fellowships towards the creation of science technology innovation,
 Grant No.~\mbox{JPMJFS2125}.
We acknowledge support from
the Fugaku Supercomputing Project (Nos.~JPMXP1020230325 and JPMXP1020230327) and 
the Data-Driven Material Research Project (No.~\mbox{JPMXP1122714694})
from the
Ministry of Education, Culture, Sports, Science, and Technology.
The numerical calculations were performed at Research Center of
Computational Science, Okazaki Research Facilities, National Institutes
of Natural Sciences (Project: 23-IMS-C052) and at the Cybermedia Center, Osaka University.

\end{acknowledgements}

%\bibliography{gkm}

\begin{thebibliography}{78}%
\makeatletter
\providecommand \@ifxundefined [1]{%
 \@ifx{#1\undefined}
}%
\providecommand \@ifnum [1]{%
 \ifnum #1\expandafter \@firstoftwo
 \else \expandafter \@secondoftwo
 \fi
}%
\providecommand \@ifx [1]{%
 \ifx #1\expandafter \@firstoftwo
 \else \expandafter \@secondoftwo
 \fi
}%
\providecommand \natexlab [1]{#1}%
\providecommand \enquote  [1]{``#1''}%
\providecommand \bibnamefont  [1]{#1}%
\providecommand \bibfnamefont [1]{#1}%
\providecommand \citenamefont [1]{#1}%
\providecommand \href@noop [0]{\@secondoftwo}%
\providecommand \href [0]{\begingroup \@sanitize@url \@href}%
\providecommand \@href[1]{\@@startlink{#1}\@@href}%
\providecommand \@@href[1]{\endgroup#1\@@endlink}%
\providecommand \@sanitize@url [0]{\catcode `\\12\catcode `\$12\catcode
  `\&12\catcode `\#12\catcode `\^12\catcode `\_12\catcode `\%12\relax}%
\providecommand \@@startlink[1]{}%
\providecommand \@@endlink[0]{}%
\providecommand \url  [0]{\begingroup\@sanitize@url \@url }%
\providecommand \@url [1]{\endgroup\@href {#1}{\urlprefix }}%
\providecommand \urlprefix  [0]{URL }%
\providecommand \Eprint [0]{\href }%
\providecommand \doibase [0]{https://doi.org/}%
\providecommand \selectlanguage [0]{\@gobble}%
\providecommand \bibinfo  [0]{\@secondoftwo}%
\providecommand \bibfield  [0]{\@secondoftwo}%
\providecommand \translation [1]{[#1]}%
\providecommand \BibitemOpen [0]{}%
\providecommand \bibitemStop [0]{}%
\providecommand \bibitemNoStop [0]{.\EOS\space}%
\providecommand \EOS [0]{\spacefactor3000\relax}%
\providecommand \BibitemShut  [1]{\csname bibitem#1\endcsname}%
\let\auto@bib@innerbib\@empty
%</preamble>
\bibitem [{\citenamefont {de~Gennes}(1979)}]{gennes1979Scaling}%
  \BibitemOpen
  \bibfield  {author} {\bibinfo {author} {\bibfnamefont {P.~G.}\ \bibnamefont
  {de~Gennes}},\ }\href@noop {} {\emph {\bibinfo {title} {Scaling Concepts in
  Polymer Physics}}}\ (\bibinfo  {publisher} {{Cornell University Press}},\
  \bibinfo {year} {1979})\BibitemShut {NoStop}%
\bibitem [{\citenamefont {Doi}\ and\ \citenamefont
  {Edwards}(1986)}]{doi1986Theory}%
  \BibitemOpen
  \bibfield  {author} {\bibinfo {author} {\bibfnamefont {M.}~\bibnamefont
  {Doi}}\ and\ \bibinfo {author} {\bibfnamefont {S.~F.}\ \bibnamefont
  {Edwards}},\ }\href@noop {} {\emph {\bibinfo {title} {The Theory of Polymer
  Dynamics}}},\ International Series of Monographs on Physics\ (\bibinfo
  {publisher} {{Oxford University Press}},\ \bibinfo {year} {1986})\BibitemShut
  {NoStop}%
\bibitem [{\citenamefont {Cates}\ and\ \citenamefont
  {Deutsch}(1986)}]{cates1986Conjectures}%
  \BibitemOpen
  \bibfield  {author} {\bibinfo {author} {\bibfnamefont {M.}~\bibnamefont
  {Cates}}\ and\ \bibinfo {author} {\bibfnamefont {J.}~\bibnamefont
  {Deutsch}},\ }\bibfield  {title} {\bibinfo {title} {Conjectures on the
  statistics of ring polymers},\ }\href
  {https://doi.org/10.1051/jphys:0198600470120212100} {\bibfield  {journal}
  {\bibinfo  {journal} {J. Phys. France}\ }\textbf {\bibinfo {volume} {47}},\
  \bibinfo {pages} {2121} (\bibinfo {year} {1986})}\BibitemShut {NoStop}%
\bibitem [{\citenamefont {Grosberg}\ \emph {et~al.}(1993)\citenamefont
  {Grosberg}, \citenamefont {Rabin}, \citenamefont {Havlin},\ and\
  \citenamefont {Neer}}]{grosberg1993Crumpled}%
  \BibitemOpen
  \bibfield  {author} {\bibinfo {author} {\bibfnamefont {A.}~\bibnamefont
  {Grosberg}}, \bibinfo {author} {\bibfnamefont {Y.}~\bibnamefont {Rabin}},
  \bibinfo {author} {\bibfnamefont {S.}~\bibnamefont {Havlin}},\ and\ \bibinfo
  {author} {\bibfnamefont {A.}~\bibnamefont {Neer}},\ }\bibfield  {title}
  {\bibinfo {title} {Crumpled {{Globule Model}} of the {{Three-Dimensional
  Structure}} of {{DNA}}},\ }\href {https://doi.org/10.1209/0295-5075/23/5/012}
  {\bibfield  {journal} {\bibinfo  {journal} {Europhys. Lett.}\ }\textbf
  {\bibinfo {volume} {23}},\ \bibinfo {pages} {373} (\bibinfo {year}
  {1993})}\BibitemShut {NoStop}%
\bibitem [{\citenamefont {Sakaue}(2011)}]{sakaue2011Ring}%
  \BibitemOpen
  \bibfield  {author} {\bibinfo {author} {\bibfnamefont {T.}~\bibnamefont
  {Sakaue}},\ }\bibfield  {title} {\bibinfo {title} {Ring {{Polymers}} in
  {{Melts}} and {{Solutions}}: {{Scaling}} and {{Crossover}}},\ }\href
  {https://doi.org/10.1103/PhysRevLett.106.167802} {\bibfield  {journal}
  {\bibinfo  {journal} {Phys. Rev. Lett.}\ }\textbf {\bibinfo {volume} {106}},\
  \bibinfo {pages} {167802} (\bibinfo {year} {2011})}\BibitemShut {NoStop}%
\bibitem [{\citenamefont {Halverson}\ \emph {et~al.}(2014)\citenamefont
  {Halverson}, \citenamefont {Smrek}, \citenamefont {Kremer},\ and\
  \citenamefont {Grosberg}}]{halverson2014Melta}%
  \BibitemOpen
  \bibfield  {author} {\bibinfo {author} {\bibfnamefont {J.~D.}\ \bibnamefont
  {Halverson}}, \bibinfo {author} {\bibfnamefont {J.}~\bibnamefont {Smrek}},
  \bibinfo {author} {\bibfnamefont {K.}~\bibnamefont {Kremer}},\ and\ \bibinfo
  {author} {\bibfnamefont {A.~Y.}\ \bibnamefont {Grosberg}},\ }\bibfield
  {title} {\bibinfo {title} {From a melt of rings to chromosome territories:
  The role of topological constraints in genome folding},\ }\href
  {https://doi.org/10.1088/0034-4885/77/2/022601} {\bibfield  {journal}
  {\bibinfo  {journal} {Rep. Prog. Phys.}\ }\textbf {\bibinfo {volume} {77}},\
  \bibinfo {pages} {022601} (\bibinfo {year} {2014})}\BibitemShut {NoStop}%
\bibitem [{\citenamefont {Ge}\ \emph {et~al.}(2016)\citenamefont {Ge},
  \citenamefont {Panyukov},\ and\ \citenamefont
  {Rubinstein}}]{ge2016SelfSimilar}%
  \BibitemOpen
  \bibfield  {author} {\bibinfo {author} {\bibfnamefont {T.}~\bibnamefont
  {Ge}}, \bibinfo {author} {\bibfnamefont {S.}~\bibnamefont {Panyukov}},\ and\
  \bibinfo {author} {\bibfnamefont {M.}~\bibnamefont {Rubinstein}},\ }\bibfield
   {title} {\bibinfo {title} {Self-{{Similar Conformations}} and {{Dynamics}}
  in {{Entangled Melts}} and {{Solutions}} of {{Nonconcatenated Ring
  Polymers}}},\ }\href {https://doi.org/10.1021/acs.macromol.5b02319}
  {\bibfield  {journal} {\bibinfo  {journal} {Macromolecules}\ }\textbf
  {\bibinfo {volume} {49}},\ \bibinfo {pages} {708} (\bibinfo {year}
  {2016})}\BibitemShut {NoStop}%
\bibitem [{\citenamefont {Kim}\ \emph {et~al.}(2021)\citenamefont {Kim},
  \citenamefont {Kim},\ and\ \citenamefont {Baig}}]{kim2021Intrinsic}%
  \BibitemOpen
  \bibfield  {author} {\bibinfo {author} {\bibfnamefont {J.}~\bibnamefont
  {Kim}}, \bibinfo {author} {\bibfnamefont {J.~M.}\ \bibnamefont {Kim}},\ and\
  \bibinfo {author} {\bibfnamefont {C.}~\bibnamefont {Baig}},\ }\bibfield
  {title} {\bibinfo {title} {Intrinsic structure and dynamics of monolayer ring
  polymer melts},\ }\href {https://doi.org/10.1039/D1SM01192H} {\bibfield
  {journal} {\bibinfo  {journal} {Soft Matter}\ }\textbf {\bibinfo {volume}
  {17}},\ \bibinfo {pages} {10703} (\bibinfo {year} {2021})}\BibitemShut
  {NoStop}%
\bibitem [{\citenamefont {M{\"u}ller}\ \emph {et~al.}(1996)\citenamefont
  {M{\"u}ller}, \citenamefont {Wittmer},\ and\ \citenamefont
  {Cates}}]{muller1996Topological}%
  \BibitemOpen
  \bibfield  {author} {\bibinfo {author} {\bibfnamefont {M.}~\bibnamefont
  {M{\"u}ller}}, \bibinfo {author} {\bibfnamefont {J.~P.}\ \bibnamefont
  {Wittmer}},\ and\ \bibinfo {author} {\bibfnamefont {M.~E.}\ \bibnamefont
  {Cates}},\ }\bibfield  {title} {\bibinfo {title} {Topological effects in ring
  polymers: {{A}} computer simulation study},\ }\href
  {https://doi.org/10.1103/PhysRevE.53.5063} {\bibfield  {journal} {\bibinfo
  {journal} {Phys. Rev. E}\ }\textbf {\bibinfo {volume} {53}},\ \bibinfo
  {pages} {5063} (\bibinfo {year} {1996})}\BibitemShut {NoStop}%
\bibitem [{\citenamefont {Smrek}\ and\ \citenamefont
  {Grosberg}(2016)}]{smrek2016Minimal}%
  \BibitemOpen
  \bibfield  {author} {\bibinfo {author} {\bibfnamefont {J.}~\bibnamefont
  {Smrek}}\ and\ \bibinfo {author} {\bibfnamefont {A.~Y.}\ \bibnamefont
  {Grosberg}},\ }\bibfield  {title} {\bibinfo {title} {Minimal {{Surfaces}} on
  {{Unconcatenated Polymer Rings}} in {{Melt}}},\ }\href
  {https://doi.org/10.1021/acsmacrolett.6b00289} {\bibfield  {journal}
  {\bibinfo  {journal} {ACS Macro Lett.}\ }\textbf {\bibinfo {volume} {5}},\
  \bibinfo {pages} {750} (\bibinfo {year} {2016})}\BibitemShut {NoStop}%
\bibitem [{\citenamefont {Landuzzi}\ \emph {et~al.}(2020)\citenamefont
  {Landuzzi}, \citenamefont {Nakamura}, \citenamefont {Michieletto},\ and\
  \citenamefont {Sakaue}}]{landuzzi2020Persistence}%
  \BibitemOpen
  \bibfield  {author} {\bibinfo {author} {\bibfnamefont {F.}~\bibnamefont
  {Landuzzi}}, \bibinfo {author} {\bibfnamefont {T.}~\bibnamefont {Nakamura}},
  \bibinfo {author} {\bibfnamefont {D.}~\bibnamefont {Michieletto}},\ and\
  \bibinfo {author} {\bibfnamefont {T.}~\bibnamefont {Sakaue}},\ }\bibfield
  {title} {\bibinfo {title} {Persistence homology of entangled rings},\ }\href
  {https://doi.org/10.1103/PhysRevResearch.2.033529} {\bibfield  {journal}
  {\bibinfo  {journal} {Phys. Rev. Research}\ }\textbf {\bibinfo {volume}
  {2}},\ \bibinfo {pages} {033529} (\bibinfo {year} {2020})}\BibitemShut
  {NoStop}%
\bibitem [{\citenamefont {Sta{\v n}o}\ \emph {et~al.}(2022)\citenamefont
  {Sta{\v n}o}, \citenamefont {Likos},\ and\ \citenamefont
  {Smrek}}]{stano2022Thread}%
  \BibitemOpen
  \bibfield  {author} {\bibinfo {author} {\bibfnamefont {R.}~\bibnamefont
  {Sta{\v n}o}}, \bibinfo {author} {\bibfnamefont {C.~N.}\ \bibnamefont
  {Likos}},\ and\ \bibinfo {author} {\bibfnamefont {J.}~\bibnamefont {Smrek}},\
  }\bibfield  {title} {\bibinfo {title} {To thread or not to thread?
  {{Effective}} potentials and threading interactions between asymmetric ring
  polymers},\ }\href {https://doi.org/10.1039/D2SM01177H} {\bibfield  {journal}
  {\bibinfo  {journal} {Soft Matter}\ }\textbf {\bibinfo {volume} {19}},\
  \bibinfo {pages} {17} (\bibinfo {year} {2022})}\BibitemShut {NoStop}%
\bibitem [{\citenamefont {Donth}(2001)}]{donth2001Glass}%
  \BibitemOpen
  \bibfield  {author} {\bibinfo {author} {\bibfnamefont {E.-J.}\ \bibnamefont
  {Donth}},\ }\href {https://doi.org/10.1007/978-3-662-04365-3} {\emph
  {\bibinfo {title} {The {{Glass Transition}}: {{Relaxation Dynamics}} in
  {{Liquids}} and {{Disordered Materials}}}}}\ (\bibinfo  {publisher}
  {{Springer Berlin Heidelberg}},\ \bibinfo {address} {{Berlin, Heidelberg}},\
  \bibinfo {year} {2001})\BibitemShut {NoStop}%
\bibitem [{\citenamefont {Lo}\ and\ \citenamefont
  {Turner}(2013)}]{lo2013Topological}%
  \BibitemOpen
  \bibfield  {author} {\bibinfo {author} {\bibfnamefont {W.-C.}\ \bibnamefont
  {Lo}}\ and\ \bibinfo {author} {\bibfnamefont {M.~S.}\ \bibnamefont
  {Turner}},\ }\bibfield  {title} {\bibinfo {title} {The topological glass in
  ring polymers},\ }\href {https://doi.org/10.1209/0295-5075/102/58005}
  {\bibfield  {journal} {\bibinfo  {journal} {EPL}\ }\textbf {\bibinfo {volume}
  {102}},\ \bibinfo {pages} {58005} (\bibinfo {year} {2013})}\BibitemShut
  {NoStop}%
\bibitem [{\citenamefont {Lee}\ \emph {et~al.}(2015)\citenamefont {Lee},
  \citenamefont {Kim},\ and\ \citenamefont {Jung}}]{lee2015Slowing}%
  \BibitemOpen
  \bibfield  {author} {\bibinfo {author} {\bibfnamefont {E.}~\bibnamefont
  {Lee}}, \bibinfo {author} {\bibfnamefont {S.}~\bibnamefont {Kim}},\ and\
  \bibinfo {author} {\bibfnamefont {Y.}~\bibnamefont {Jung}},\ }\bibfield
  {title} {\bibinfo {title} {Slowing {{Down}} of {{Ring Polymer Diffusion
  Caused}} by {{Inter-Ring Threading}}},\ }\href
  {https://doi.org/10.1002/marc.201400713} {\bibfield  {journal} {\bibinfo
  {journal} {Macromol. Rapid Commun.}\ }\textbf {\bibinfo {volume} {36}},\
  \bibinfo {pages} {1115} (\bibinfo {year} {2015})}\BibitemShut {NoStop}%
\bibitem [{\citenamefont {Michieletto}\ and\ \citenamefont
  {Turner}(2016)}]{michieletto2016Topologically}%
  \BibitemOpen
  \bibfield  {author} {\bibinfo {author} {\bibfnamefont {D.}~\bibnamefont
  {Michieletto}}\ and\ \bibinfo {author} {\bibfnamefont {M.~S.}\ \bibnamefont
  {Turner}},\ }\bibfield  {title} {\bibinfo {title} {A topologically driven
  glass in ring polymers},\ }\href {https://doi.org/10.1073/pnas.1520665113}
  {\bibfield  {journal} {\bibinfo  {journal} {Proc. Natl. Acad. Sci. U.S.A.}\
  }\textbf {\bibinfo {volume} {113}},\ \bibinfo {pages} {5195} (\bibinfo {year}
  {2016})}\BibitemShut {NoStop}%
\bibitem [{\citenamefont {Michieletto}\ \emph
  {et~al.}(2017{\natexlab{a}})\citenamefont {Michieletto}, \citenamefont
  {Nahali},\ and\ \citenamefont {Rosa}}]{michieletto2017Glassiness}%
  \BibitemOpen
  \bibfield  {author} {\bibinfo {author} {\bibfnamefont {D.}~\bibnamefont
  {Michieletto}}, \bibinfo {author} {\bibfnamefont {N.}~\bibnamefont
  {Nahali}},\ and\ \bibinfo {author} {\bibfnamefont {A.}~\bibnamefont {Rosa}},\
  }\bibfield  {title} {\bibinfo {title} {Glassiness and {{Heterogeneous
  Dynamics}} in {{Dense Solutions}} of {{Ring Polymers}}},\ }\href
  {https://doi.org/10.1103/PhysRevLett.119.197801} {\bibfield  {journal}
  {\bibinfo  {journal} {Phys. Rev. Lett.}\ }\textbf {\bibinfo {volume} {119}},\
  \bibinfo {pages} {197801} (\bibinfo {year} {2017}{\natexlab{a}})}\BibitemShut
  {NoStop}%
\bibitem [{\citenamefont {Michieletto}\ \emph
  {et~al.}(2017{\natexlab{b}})\citenamefont {Michieletto}, \citenamefont
  {Marenduzzo}, \citenamefont {Orlandini},\ and\ \citenamefont
  {Turner}}]{michieletto2017Ring}%
  \BibitemOpen
  \bibfield  {author} {\bibinfo {author} {\bibfnamefont {D.}~\bibnamefont
  {Michieletto}}, \bibinfo {author} {\bibfnamefont {D.}~\bibnamefont
  {Marenduzzo}}, \bibinfo {author} {\bibfnamefont {E.}~\bibnamefont
  {Orlandini}},\ and\ \bibinfo {author} {\bibfnamefont {M.}~\bibnamefont
  {Turner}},\ }\bibfield  {title} {\bibinfo {title} {Ring {{Polymers}}:
  {{Threadings}}, {{Knot Electrophoresis}} and {{Topological Glasses}}},\
  }\href {https://doi.org/10.3390/polym9080349} {\bibfield  {journal} {\bibinfo
   {journal} {Polymers}\ }\textbf {\bibinfo {volume} {9}},\ \bibinfo {pages}
  {349} (\bibinfo {year} {2017}{\natexlab{b}})}\BibitemShut {NoStop}%
\bibitem [{\citenamefont {Sakaue}(2018)}]{sakaue2018Topological}%
  \BibitemOpen
  \bibfield  {author} {\bibinfo {author} {\bibfnamefont {T.}~\bibnamefont
  {Sakaue}},\ }\bibfield  {title} {\bibinfo {title} {Topological free volume
  and quasi-glassy dynamics in the melt of ring polymers},\ }\href
  {https://doi.org/10.1039/C8SM00968F} {\bibfield  {journal} {\bibinfo
  {journal} {Soft Matter}\ }\textbf {\bibinfo {volume} {14}},\ \bibinfo {pages}
  {7507} (\bibinfo {year} {2018})}\BibitemShut {NoStop}%
\bibitem [{\citenamefont {G{\'o}mez}\ \emph {et~al.}(2020)\citenamefont
  {G{\'o}mez}, \citenamefont {Garc{\'i}a},\ and\ \citenamefont
  {P{\"o}schel}}]{gomez2020Packinga}%
  \BibitemOpen
  \bibfield  {author} {\bibinfo {author} {\bibfnamefont {L.~R.}\ \bibnamefont
  {G{\'o}mez}}, \bibinfo {author} {\bibfnamefont {N.~A.}\ \bibnamefont
  {Garc{\'i}a}},\ and\ \bibinfo {author} {\bibfnamefont {T.}~\bibnamefont
  {P{\"o}schel}},\ }\bibfield  {title} {\bibinfo {title} {Packing structure of
  semiflexible rings},\ }\href {https://doi.org/10.1073/pnas.1914268117}
  {\bibfield  {journal} {\bibinfo  {journal} {Proc. Natl. Acad. Sci. U.S.A.}\
  }\textbf {\bibinfo {volume} {117}},\ \bibinfo {pages} {3382} (\bibinfo {year}
  {2020})}\BibitemShut {NoStop}%
\bibitem [{\citenamefont {Smrek}\ \emph {et~al.}(2020)\citenamefont {Smrek},
  \citenamefont {Chubak}, \citenamefont {Likos},\ and\ \citenamefont
  {Kremer}}]{smrek2020Active}%
  \BibitemOpen
  \bibfield  {author} {\bibinfo {author} {\bibfnamefont {J.}~\bibnamefont
  {Smrek}}, \bibinfo {author} {\bibfnamefont {I.}~\bibnamefont {Chubak}},
  \bibinfo {author} {\bibfnamefont {C.~N.}\ \bibnamefont {Likos}},\ and\
  \bibinfo {author} {\bibfnamefont {K.}~\bibnamefont {Kremer}},\ }\bibfield
  {title} {\bibinfo {title} {Active topological glass},\ }\href
  {https://doi.org/10.1038/s41467-019-13696-z} {\bibfield  {journal} {\bibinfo
  {journal} {Nat. Commun.}\ }\textbf {\bibinfo {volume} {11}},\ \bibinfo
  {pages} {26} (\bibinfo {year} {2020})}\BibitemShut {NoStop}%
\bibitem [{\citenamefont {Chubak}\ \emph {et~al.}(2020)\citenamefont {Chubak},
  \citenamefont {Likos}, \citenamefont {Kremer},\ and\ \citenamefont
  {Smrek}}]{chubak2020Emergence}%
  \BibitemOpen
  \bibfield  {author} {\bibinfo {author} {\bibfnamefont {I.}~\bibnamefont
  {Chubak}}, \bibinfo {author} {\bibfnamefont {C.~N.}\ \bibnamefont {Likos}},
  \bibinfo {author} {\bibfnamefont {K.}~\bibnamefont {Kremer}},\ and\ \bibinfo
  {author} {\bibfnamefont {J.}~\bibnamefont {Smrek}},\ }\bibfield  {title}
  {\bibinfo {title} {Emergence of active topological glass through directed
  chain dynamics and nonequilibrium phase segregation},\ }\href
  {https://doi.org/10.1103/PhysRevResearch.2.043249} {\bibfield  {journal}
  {\bibinfo  {journal} {Phys. Rev. Research}\ }\textbf {\bibinfo {volume}
  {2}},\ \bibinfo {pages} {043249} (\bibinfo {year} {2020})}\BibitemShut
  {NoStop}%
\bibitem [{\citenamefont {Michieletto}\ and\ \citenamefont
  {Sakaue}(2021)}]{michieletto2021Dynamical}%
  \BibitemOpen
  \bibfield  {author} {\bibinfo {author} {\bibfnamefont {D.}~\bibnamefont
  {Michieletto}}\ and\ \bibinfo {author} {\bibfnamefont {T.}~\bibnamefont
  {Sakaue}},\ }\bibfield  {title} {\bibinfo {title} {Dynamical {{Entanglement}}
  and {{Cooperative Dynamics}} in {{Entangled Solutions}} of {{Ring}} and
  {{Linear Polymers}}},\ }\href {https://doi.org/10.1021/acsmacrolett.0c00551}
  {\bibfield  {journal} {\bibinfo  {journal} {ACS Macro Lett.}\ }\textbf
  {\bibinfo {volume} {10}},\ \bibinfo {pages} {129} (\bibinfo {year}
  {2021})}\BibitemShut {NoStop}%
\bibitem [{\citenamefont {Chubak}\ \emph {et~al.}(2022)\citenamefont {Chubak},
  \citenamefont {Pachong}, \citenamefont {Kremer}, \citenamefont {Likos},\ and\
  \citenamefont {Smrek}}]{chubak2022Active}%
  \BibitemOpen
  \bibfield  {author} {\bibinfo {author} {\bibfnamefont {I.}~\bibnamefont
  {Chubak}}, \bibinfo {author} {\bibfnamefont {S.~M.}\ \bibnamefont {Pachong}},
  \bibinfo {author} {\bibfnamefont {K.}~\bibnamefont {Kremer}}, \bibinfo
  {author} {\bibfnamefont {C.~N.}\ \bibnamefont {Likos}},\ and\ \bibinfo
  {author} {\bibfnamefont {J.}~\bibnamefont {Smrek}},\ }\bibfield  {title}
  {\bibinfo {title} {Active {{Topological Glass Confined}} within a {{Spherical
  Cavity}}},\ }\href {https://doi.org/10.1021/acs.macromol.1c02471} {\bibfield
  {journal} {\bibinfo  {journal} {Macromolecules}\ }\textbf {\bibinfo {volume}
  {55}},\ \bibinfo {pages} {956} (\bibinfo {year} {2022})}\BibitemShut
  {NoStop}%
\bibitem [{\citenamefont {B{\"o}hmer}\ \emph {et~al.}(1996)\citenamefont
  {B{\"o}hmer}, \citenamefont {Hinze}, \citenamefont {Diezemann}, \citenamefont
  {Geil},\ and\ \citenamefont {Sillescu}}]{bohmer1996Dynamic}%
  \BibitemOpen
  \bibfield  {author} {\bibinfo {author} {\bibfnamefont {R.}~\bibnamefont
  {B{\"o}hmer}}, \bibinfo {author} {\bibfnamefont {G.}~\bibnamefont {Hinze}},
  \bibinfo {author} {\bibfnamefont {G.}~\bibnamefont {Diezemann}}, \bibinfo
  {author} {\bibfnamefont {B.}~\bibnamefont {Geil}},\ and\ \bibinfo {author}
  {\bibfnamefont {H.}~\bibnamefont {Sillescu}},\ }\bibfield  {title} {\bibinfo
  {title} {Dynamic heterogeneity in supercooled ortho-terphenyl studied by
  multidimensional deuteron {{NMR}}},\ }\href
  {https://doi.org/10.1209/epl/i1996-00186-5} {\bibfield  {journal} {\bibinfo
  {journal} {Europhys. Lett.}\ }\textbf {\bibinfo {volume} {36}},\ \bibinfo
  {pages} {55} (\bibinfo {year} {1996})}\BibitemShut {NoStop}%
\bibitem [{\citenamefont {Richert}\ and\ \citenamefont
  {Richert}(1998)}]{richert1998Dynamic}%
  \BibitemOpen
  \bibfield  {author} {\bibinfo {author} {\bibfnamefont {R.}~\bibnamefont
  {Richert}}\ and\ \bibinfo {author} {\bibfnamefont {M.}~\bibnamefont
  {Richert}},\ }\bibfield  {title} {\bibinfo {title} {Dynamic heterogeneity,
  spatially distributed stretched-exponential patterns, and transient
  dispersions in solvation dynamics},\ }\href
  {https://doi.org/10.1103/PhysRevE.58.779} {\bibfield  {journal} {\bibinfo
  {journal} {Phys. Rev. E}\ }\textbf {\bibinfo {volume} {58}},\ \bibinfo
  {pages} {779} (\bibinfo {year} {1998})}\BibitemShut {NoStop}%
\bibitem [{\citenamefont {Ediger}(2000)}]{ediger2000Spatially}%
  \BibitemOpen
  \bibfield  {author} {\bibinfo {author} {\bibfnamefont {M.~D.}\ \bibnamefont
  {Ediger}},\ }\bibfield  {title} {\bibinfo {title} {Spatially {{Heterogeneous
  Dynamics}} in {{Supercooled Liquids}}},\ }\href
  {https://doi.org/10.1146/annurev.physchem.51.1.99} {\bibfield  {journal}
  {\bibinfo  {journal} {Annu. Rev. Phys. Chem.}\ }\textbf {\bibinfo {volume}
  {51}},\ \bibinfo {pages} {99} (\bibinfo {year} {2000})}\BibitemShut {NoStop}%
\bibitem [{\citenamefont {Hurley}\ and\ \citenamefont
  {Harrowell}(1995)}]{hurley1995Kinetic}%
  \BibitemOpen
  \bibfield  {author} {\bibinfo {author} {\bibfnamefont {M.~M.}\ \bibnamefont
  {Hurley}}\ and\ \bibinfo {author} {\bibfnamefont {P.}~\bibnamefont
  {Harrowell}},\ }\bibfield  {title} {\bibinfo {title} {Kinetic structure of a
  two-dimensional liquid},\ }\href {https://doi.org/10.1103/PhysRevE.52.1694}
  {\bibfield  {journal} {\bibinfo  {journal} {Phys. Rev. E}\ }\textbf {\bibinfo
  {volume} {52}},\ \bibinfo {pages} {1694} (\bibinfo {year}
  {1995})}\BibitemShut {NoStop}%
\bibitem [{\citenamefont {Kob}\ \emph {et~al.}(1997)\citenamefont {Kob},
  \citenamefont {Donati}, \citenamefont {Plimpton}, \citenamefont {Poole},\
  and\ \citenamefont {Glotzer}}]{kob1997Dynamical}%
  \BibitemOpen
  \bibfield  {author} {\bibinfo {author} {\bibfnamefont {W.}~\bibnamefont
  {Kob}}, \bibinfo {author} {\bibfnamefont {C.}~\bibnamefont {Donati}},
  \bibinfo {author} {\bibfnamefont {S.~J.}\ \bibnamefont {Plimpton}}, \bibinfo
  {author} {\bibfnamefont {P.~H.}\ \bibnamefont {Poole}},\ and\ \bibinfo
  {author} {\bibfnamefont {S.~C.}\ \bibnamefont {Glotzer}},\ }\bibfield
  {title} {\bibinfo {title} {Dynamical {{Heterogeneities}} in a {{Supercooled
  Lennard-Jones Liquid}}},\ }\href
  {https://doi.org/10.1103/PhysRevLett.79.2827} {\bibfield  {journal} {\bibinfo
   {journal} {Phys. Rev. Lett.}\ }\textbf {\bibinfo {volume} {79}},\ \bibinfo
  {pages} {2827} (\bibinfo {year} {1997})}\BibitemShut {NoStop}%
\bibitem [{\citenamefont {Yamamoto}\ and\ \citenamefont
  {Onuki}(1997)}]{yamamoto1997Kinetic}%
  \BibitemOpen
  \bibfield  {author} {\bibinfo {author} {\bibfnamefont {R.}~\bibnamefont
  {Yamamoto}}\ and\ \bibinfo {author} {\bibfnamefont {A.}~\bibnamefont
  {Onuki}},\ }\bibfield  {title} {\bibinfo {title} {Kinetic {{Heterogeneities}}
  in a {{Highly Supercooled Liquid}}},\ }\href
  {https://doi.org/10.1143/JPSJ.66.2545} {\bibfield  {journal} {\bibinfo
  {journal} {J. Phys. Soc. Jpn.}\ }\textbf {\bibinfo {volume} {66}},\ \bibinfo
  {pages} {2545} (\bibinfo {year} {1997})}\BibitemShut {NoStop}%
\bibitem [{\citenamefont {Donati}\ \emph {et~al.}(1998)\citenamefont {Donati},
  \citenamefont {Douglas}, \citenamefont {Kob}, \citenamefont {Plimpton},
  \citenamefont {Poole},\ and\ \citenamefont {Glotzer}}]{donati1998Stringlike}%
  \BibitemOpen
  \bibfield  {author} {\bibinfo {author} {\bibfnamefont {C.}~\bibnamefont
  {Donati}}, \bibinfo {author} {\bibfnamefont {J.~F.}\ \bibnamefont {Douglas}},
  \bibinfo {author} {\bibfnamefont {W.}~\bibnamefont {Kob}}, \bibinfo {author}
  {\bibfnamefont {S.~J.}\ \bibnamefont {Plimpton}}, \bibinfo {author}
  {\bibfnamefont {P.~H.}\ \bibnamefont {Poole}},\ and\ \bibinfo {author}
  {\bibfnamefont {S.~C.}\ \bibnamefont {Glotzer}},\ }\bibfield  {title}
  {\bibinfo {title} {Stringlike {{Cooperative Motion}} in a {{Supercooled
  Liquid}}},\ }\href {https://doi.org/10.1103/PhysRevLett.80.2338} {\bibfield
  {journal} {\bibinfo  {journal} {Phys. Rev. Lett.}\ }\textbf {\bibinfo
  {volume} {80}},\ \bibinfo {pages} {2338} (\bibinfo {year}
  {1998})}\BibitemShut {NoStop}%
\bibitem [{\citenamefont {Hurley}\ and\ \citenamefont
  {Harrowell}(1996)}]{hurley1996Non}%
  \BibitemOpen
  \bibfield  {author} {\bibinfo {author} {\bibfnamefont {M.~M.}\ \bibnamefont
  {Hurley}}\ and\ \bibinfo {author} {\bibfnamefont {P.}~\bibnamefont
  {Harrowell}},\ }\bibfield  {title} {\bibinfo {title} {Non-{{Gaussian}}
  behavior and the dynamical complexity of particle motion in a dense
  two-dimensional liquid},\ }\href {https://doi.org/10.1063/1.472941}
  {\bibfield  {journal} {\bibinfo  {journal} {J. Chem. Phys.}\ }\textbf
  {\bibinfo {volume} {105}},\ \bibinfo {pages} {10521} (\bibinfo {year}
  {1996})}\BibitemShut {NoStop}%
\bibitem [{\citenamefont {Shell}\ \emph {et~al.}(2005)\citenamefont {Shell},
  \citenamefont {Debenedetti},\ and\ \citenamefont
  {Stillinger}}]{shell2005Dynamic}%
  \BibitemOpen
  \bibfield  {author} {\bibinfo {author} {\bibfnamefont {M.~S.}\ \bibnamefont
  {Shell}}, \bibinfo {author} {\bibfnamefont {P.~G.}\ \bibnamefont
  {Debenedetti}},\ and\ \bibinfo {author} {\bibfnamefont {F.~H.}\ \bibnamefont
  {Stillinger}},\ }\bibfield  {title} {\bibinfo {title} {Dynamic heterogeneity
  and non-{{Gaussian}} behaviour in a model supercooled liquid},\ }\href
  {https://doi.org/10.1088/0953-8984/17/49/002} {\bibfield  {journal} {\bibinfo
   {journal} {J. Phys.: Condens. Matter}\ }\textbf {\bibinfo {volume} {17}},\
  \bibinfo {pages} {S4035} (\bibinfo {year} {2005})}\BibitemShut {NoStop}%
\bibitem [{\citenamefont {Flenner}\ and\ \citenamefont
  {Szamel}(2005)}]{flenner2005Relaxation}%
  \BibitemOpen
  \bibfield  {author} {\bibinfo {author} {\bibfnamefont {E.}~\bibnamefont
  {Flenner}}\ and\ \bibinfo {author} {\bibfnamefont {G.}~\bibnamefont
  {Szamel}},\ }\bibfield  {title} {\bibinfo {title} {Relaxation in a glassy
  binary mixture: {{Comparison}} of the mode-coupling theory to a {{Brownian}}
  dynamics simulation},\ }\href {https://doi.org/10.1103/PhysRevE.72.031508}
  {\bibfield  {journal} {\bibinfo  {journal} {Phys. Rev. E}\ }\textbf {\bibinfo
  {volume} {72}},\ \bibinfo {pages} {031508} (\bibinfo {year}
  {2005})}\BibitemShut {NoStop}%
\bibitem [{\citenamefont {Saltzman}\ and\ \citenamefont
  {Schweizer}(2006)}]{saltzman2006NonGaussian}%
  \BibitemOpen
  \bibfield  {author} {\bibinfo {author} {\bibfnamefont {E.~J.}\ \bibnamefont
  {Saltzman}}\ and\ \bibinfo {author} {\bibfnamefont {K.~S.}\ \bibnamefont
  {Schweizer}},\ }\bibfield  {title} {\bibinfo {title} {Non-{{Gaussian}}
  effects, space-time decoupling, and mobility bifurcation in glassy
  hard-sphere fluids and suspensions},\ }\href
  {https://doi.org/10.1103/PhysRevE.74.061501} {\bibfield  {journal} {\bibinfo
  {journal} {Phys. Rev. E}\ }\textbf {\bibinfo {volume} {74}},\ \bibinfo
  {pages} {061501} (\bibinfo {year} {2006})}\BibitemShut {NoStop}%
\bibitem [{\citenamefont {Chaudhuri}\ \emph {et~al.}(2007)\citenamefont
  {Chaudhuri}, \citenamefont {Berthier},\ and\ \citenamefont
  {Kob}}]{chaudhuri2007Universal}%
  \BibitemOpen
  \bibfield  {author} {\bibinfo {author} {\bibfnamefont {P.}~\bibnamefont
  {Chaudhuri}}, \bibinfo {author} {\bibfnamefont {L.}~\bibnamefont
  {Berthier}},\ and\ \bibinfo {author} {\bibfnamefont {W.}~\bibnamefont
  {Kob}},\ }\bibfield  {title} {\bibinfo {title} {Universal {{Nature}} of
  {{Particle Displacements}} close to {{Glass}} and {{Jamming Transitions}}},\
  }\href {https://doi.org/10.1103/PhysRevLett.99.060604} {\bibfield  {journal}
  {\bibinfo  {journal} {Phys. Rev. Lett.}\ }\textbf {\bibinfo {volume} {99}},\
  \bibinfo {pages} {060604} (\bibinfo {year} {2007})}\BibitemShut {NoStop}%
\bibitem [{\citenamefont {Aichele}\ \emph {et~al.}(2003)\citenamefont
  {Aichele}, \citenamefont {Gebremichael}, \citenamefont {Starr}, \citenamefont
  {Baschnagel},\ and\ \citenamefont {Glotzer}}]{aichele2003Polymerspecific}%
  \BibitemOpen
  \bibfield  {author} {\bibinfo {author} {\bibfnamefont {M.}~\bibnamefont
  {Aichele}}, \bibinfo {author} {\bibfnamefont {Y.}~\bibnamefont
  {Gebremichael}}, \bibinfo {author} {\bibfnamefont {F.~W.}\ \bibnamefont
  {Starr}}, \bibinfo {author} {\bibfnamefont {J.}~\bibnamefont {Baschnagel}},\
  and\ \bibinfo {author} {\bibfnamefont {S.~C.}\ \bibnamefont {Glotzer}},\
  }\bibfield  {title} {\bibinfo {title} {Polymer-specific effects of bulk
  relaxation and stringlike correlated motion in the dynamics of a supercooled
  polymer melt},\ }\href {https://doi.org/10.1063/1.1597473} {\bibfield
  {journal} {\bibinfo  {journal} {J. Chem. Phys.}\ }\textbf {\bibinfo {volume}
  {119}},\ \bibinfo {pages} {5290} (\bibinfo {year} {2003})}\BibitemShut
  {NoStop}%
\bibitem [{\citenamefont {Peter}\ \emph {et~al.}(2009)\citenamefont {Peter},
  \citenamefont {Meyer},\ and\ \citenamefont {Baschnagel}}]{peter2009MD}%
  \BibitemOpen
  \bibfield  {author} {\bibinfo {author} {\bibfnamefont {S.}~\bibnamefont
  {Peter}}, \bibinfo {author} {\bibfnamefont {H.}~\bibnamefont {Meyer}},\ and\
  \bibinfo {author} {\bibfnamefont {J.}~\bibnamefont {Baschnagel}},\ }\bibfield
   {title} {\bibinfo {title} {{{MD}} simulation of concentrated polymer
  solutions: {{Structural}} relaxation near the glass transition},\ }\href
  {https://doi.org/10.1140/epje/i2008-10372-9} {\bibfield  {journal} {\bibinfo
  {journal} {Eur. Phys. J. E}\ }\textbf {\bibinfo {volume} {28}},\ \bibinfo
  {pages} {147} (\bibinfo {year} {2009})}\BibitemShut {NoStop}%
\bibitem [{\citenamefont {Pan}\ and\ \citenamefont
  {Sun}(2018)}]{pan2018Diffusion}%
  \BibitemOpen
  \bibfield  {author} {\bibinfo {author} {\bibfnamefont {D.}~\bibnamefont
  {Pan}}\ and\ \bibinfo {author} {\bibfnamefont {Z.-Y.}\ \bibnamefont {Sun}},\
  }\bibfield  {title} {\bibinfo {title} {Diffusion and {{Relaxation Dynamics}}
  of {{Supercooled Polymer Melts}}},\ }\href
  {https://doi.org/10.1007/s10118-018-2132-9} {\bibfield  {journal} {\bibinfo
  {journal} {Chin. J. Polym. Sci.}\ }\textbf {\bibinfo {volume} {36}},\
  \bibinfo {pages} {1187} (\bibinfo {year} {2018})}\BibitemShut {NoStop}%
\bibitem [{\citenamefont {Goto}\ \emph {et~al.}(2021)\citenamefont {Goto},
  \citenamefont {Kim},\ and\ \citenamefont {Matubayasi}}]{goto2021Effects}%
  \BibitemOpen
  \bibfield  {author} {\bibinfo {author} {\bibfnamefont {S.}~\bibnamefont
  {Goto}}, \bibinfo {author} {\bibfnamefont {K.}~\bibnamefont {Kim}},\ and\
  \bibinfo {author} {\bibfnamefont {N.}~\bibnamefont {Matubayasi}},\ }\bibfield
   {title} {\bibinfo {title} {Effects of chain length on {{Rouse}} modes and
  non-{{Gaussianity}} in linear and ring polymer melts},\ }\href
  {https://doi.org/10.1063/5.0061281} {\bibfield  {journal} {\bibinfo
  {journal} {J. Chem. Phys.}\ }\textbf {\bibinfo {volume} {155}},\ \bibinfo
  {pages} {124901} (\bibinfo {year} {2021})},\ \Eprint
  {https://arxiv.org/abs/2106.13467} {arxiv:2106.13467} \BibitemShut {NoStop}%
\bibitem [{\citenamefont {Br{\'a}s}\ \emph {et~al.}(2014)\citenamefont
  {Br{\'a}s}, \citenamefont {Goo{\ss}en}, \citenamefont {Krutyeva},
  \citenamefont {Radulescu}, \citenamefont {Farago}, \citenamefont {Allgaier},
  \citenamefont {{Pyckhout-Hintzen}}, \citenamefont {Wischnewski},\ and\
  \citenamefont {Richter}}]{bras2014Compact}%
  \BibitemOpen
  \bibfield  {author} {\bibinfo {author} {\bibfnamefont {A.~R.}\ \bibnamefont
  {Br{\'a}s}}, \bibinfo {author} {\bibfnamefont {S.}~\bibnamefont
  {Goo{\ss}en}}, \bibinfo {author} {\bibfnamefont {M.}~\bibnamefont
  {Krutyeva}}, \bibinfo {author} {\bibfnamefont {A.}~\bibnamefont {Radulescu}},
  \bibinfo {author} {\bibfnamefont {B.}~\bibnamefont {Farago}}, \bibinfo
  {author} {\bibfnamefont {J.}~\bibnamefont {Allgaier}}, \bibinfo {author}
  {\bibfnamefont {W.}~\bibnamefont {{Pyckhout-Hintzen}}}, \bibinfo {author}
  {\bibfnamefont {A.}~\bibnamefont {Wischnewski}},\ and\ \bibinfo {author}
  {\bibfnamefont {D.}~\bibnamefont {Richter}},\ }\bibfield  {title} {\bibinfo
  {title} {Compact structure and non-{{Gaussian}} dynamics of ring polymer
  melts},\ }\href {https://doi.org/10.1039/C3SM52717D} {\bibfield  {journal}
  {\bibinfo  {journal} {Soft Matter}\ }\textbf {\bibinfo {volume} {10}},\
  \bibinfo {pages} {3649} (\bibinfo {year} {2014})}\BibitemShut {NoStop}%
\bibitem [{\citenamefont {Roy}\ \emph {et~al.}(2022)\citenamefont {Roy},
  \citenamefont {Chaudhuri},\ and\ \citenamefont {Vemparala}}]{roy2022Effect}%
  \BibitemOpen
  \bibfield  {author} {\bibinfo {author} {\bibfnamefont {P.~K.}\ \bibnamefont
  {Roy}}, \bibinfo {author} {\bibfnamefont {P.}~\bibnamefont {Chaudhuri}},\
  and\ \bibinfo {author} {\bibfnamefont {S.}~\bibnamefont {Vemparala}},\
  }\bibfield  {title} {\bibinfo {title} {Effect of ring stiffness and ambient
  pressure on the dynamical slowdown in ring polymers},\ }\href
  {https://doi.org/10.1039/D1SM01754C} {\bibfield  {journal} {\bibinfo
  {journal} {Soft Matter}\ }\textbf {\bibinfo {volume} {18}},\ \bibinfo {pages}
  {2959} (\bibinfo {year} {2022})}\BibitemShut {NoStop}%
\bibitem [{\citenamefont {Halverson}\ \emph
  {et~al.}(2011{\natexlab{a}})\citenamefont {Halverson}, \citenamefont {Lee},
  \citenamefont {Grest}, \citenamefont {Grosberg},\ and\ \citenamefont
  {Kremer}}]{halverson2011Molecular}%
  \BibitemOpen
  \bibfield  {author} {\bibinfo {author} {\bibfnamefont {J.~D.}\ \bibnamefont
  {Halverson}}, \bibinfo {author} {\bibfnamefont {W.~B.}\ \bibnamefont {Lee}},
  \bibinfo {author} {\bibfnamefont {G.~S.}\ \bibnamefont {Grest}}, \bibinfo
  {author} {\bibfnamefont {A.~Y.}\ \bibnamefont {Grosberg}},\ and\ \bibinfo
  {author} {\bibfnamefont {K.}~\bibnamefont {Kremer}},\ }\bibfield  {title}
  {\bibinfo {title} {Molecular dynamics simulation study of nonconcatenated
  ring polymers in a melt. {{I}}. {{Statics}}},\ }\href
  {https://doi.org/10.1063/1.3587137} {\bibfield  {journal} {\bibinfo
  {journal} {J. Chem. Phys.}\ }\textbf {\bibinfo {volume} {134}},\ \bibinfo
  {pages} {204904} (\bibinfo {year} {2011}{\natexlab{a}})}\BibitemShut
  {NoStop}%
\bibitem [{\citenamefont {Halverson}\ \emph
  {et~al.}(2011{\natexlab{b}})\citenamefont {Halverson}, \citenamefont {Lee},
  \citenamefont {Grest}, \citenamefont {Grosberg},\ and\ \citenamefont
  {Kremer}}]{halverson2011Moleculara}%
  \BibitemOpen
  \bibfield  {author} {\bibinfo {author} {\bibfnamefont {J.~D.}\ \bibnamefont
  {Halverson}}, \bibinfo {author} {\bibfnamefont {W.~B.}\ \bibnamefont {Lee}},
  \bibinfo {author} {\bibfnamefont {G.~S.}\ \bibnamefont {Grest}}, \bibinfo
  {author} {\bibfnamefont {A.~Y.}\ \bibnamefont {Grosberg}},\ and\ \bibinfo
  {author} {\bibfnamefont {K.}~\bibnamefont {Kremer}},\ }\bibfield  {title}
  {\bibinfo {title} {Molecular dynamics simulation study of nonconcatenated
  ring polymers in a melt. {{II}}. {{Dynamics}}},\ }\href
  {https://doi.org/10.1063/1.3587138} {\bibfield  {journal} {\bibinfo
  {journal} {J. Chem. Phys.}\ }\textbf {\bibinfo {volume} {134}},\ \bibinfo
  {pages} {204905} (\bibinfo {year} {2011}{\natexlab{b}})}\BibitemShut
  {NoStop}%
\bibitem [{\citenamefont {Halverson}\ \emph {et~al.}(2012)\citenamefont
  {Halverson}, \citenamefont {Grest}, \citenamefont {Grosberg},\ and\
  \citenamefont {Kremer}}]{halverson2012Rheologya}%
  \BibitemOpen
  \bibfield  {author} {\bibinfo {author} {\bibfnamefont {J.~D.}\ \bibnamefont
  {Halverson}}, \bibinfo {author} {\bibfnamefont {G.~S.}\ \bibnamefont
  {Grest}}, \bibinfo {author} {\bibfnamefont {A.~Y.}\ \bibnamefont
  {Grosberg}},\ and\ \bibinfo {author} {\bibfnamefont {K.}~\bibnamefont
  {Kremer}},\ }\bibfield  {title} {\bibinfo {title} {Rheology of {{Ring Polymer
  Melts}}: {{From Linear Contaminants}} to {{Ring-Linear Blends}}},\ }\href
  {https://doi.org/10.1103/PhysRevLett.108.038301} {\bibfield  {journal}
  {\bibinfo  {journal} {Phys. Rev. Lett.}\ }\textbf {\bibinfo {volume} {108}},\
  \bibinfo {pages} {038301} (\bibinfo {year} {2012})}\BibitemShut {NoStop}%
\bibitem [{\citenamefont {Kremer}\ and\ \citenamefont
  {Grest}(1990)}]{kremer1990Dynamics}%
  \BibitemOpen
  \bibfield  {author} {\bibinfo {author} {\bibfnamefont {K.}~\bibnamefont
  {Kremer}}\ and\ \bibinfo {author} {\bibfnamefont {G.~S.}\ \bibnamefont
  {Grest}},\ }\bibfield  {title} {\bibinfo {title} {Dynamics of entangled
  linear polymer melts: {{A}} molecular-dynamics simulation},\ }\href
  {https://doi.org/10.1063/1.458541} {\bibfield  {journal} {\bibinfo  {journal}
  {J. Chem. Phys.}\ }\textbf {\bibinfo {volume} {92}},\ \bibinfo {pages} {5057}
  (\bibinfo {year} {1990})}\BibitemShut {NoStop}%
\bibitem [{\citenamefont {Plimpton}(1995)}]{plimpton1995Fast}%
  \BibitemOpen
  \bibfield  {author} {\bibinfo {author} {\bibfnamefont {S.}~\bibnamefont
  {Plimpton}},\ }\bibfield  {title} {\bibinfo {title} {Fast {{Parallel
  Algorithms}} for {{Short-Range Molecular Dynamics}}},\ }\href
  {https://doi.org/10.1006/jcph.1995.1039} {\bibfield  {journal} {\bibinfo
  {journal} {J. Comput. Phys.}\ }\textbf {\bibinfo {volume} {117}},\ \bibinfo
  {pages} {1} (\bibinfo {year} {1995})}\BibitemShut {NoStop}%
\bibitem [{\citenamefont {Hsu}\ and\ \citenamefont
  {Kremer}(2016)}]{hsu2016Static}%
  \BibitemOpen
  \bibfield  {author} {\bibinfo {author} {\bibfnamefont {H.-P.}\ \bibnamefont
  {Hsu}}\ and\ \bibinfo {author} {\bibfnamefont {K.}~\bibnamefont {Kremer}},\
  }\bibfield  {title} {\bibinfo {title} {Static and dynamic properties of large
  polymer melts in equilibrium},\ }\href {https://doi.org/10.1063/1.4946033}
  {\bibfield  {journal} {\bibinfo  {journal} {J. Chem. Phys.}\ }\textbf
  {\bibinfo {volume} {144}},\ \bibinfo {pages} {154907} (\bibinfo {year}
  {2016})}\BibitemShut {NoStop}%
\bibitem [{\citenamefont {Hsu}\ and\ \citenamefont
  {Kremer}(2017)}]{hsu2017Detailed}%
  \BibitemOpen
  \bibfield  {author} {\bibinfo {author} {\bibfnamefont {H.-P.}\ \bibnamefont
  {Hsu}}\ and\ \bibinfo {author} {\bibfnamefont {K.}~\bibnamefont {Kremer}},\
  }\bibfield  {title} {\bibinfo {title} {Detailed analysis of {{Rouse}} mode
  and dynamic scattering function of highly entangled polymer melts in
  equilibrium},\ }\href {https://doi.org/10.1140/epjst/e2016-60322-5}
  {\bibfield  {journal} {\bibinfo  {journal} {Eur. Phys. J. Spec. Top.}\
  }\textbf {\bibinfo {volume} {226}},\ \bibinfo {pages} {693} (\bibinfo {year}
  {2017})}\BibitemShut {NoStop}%
\bibitem [{\citenamefont {Parisi}\ \emph {et~al.}(2021)\citenamefont {Parisi},
  \citenamefont {Costanzo}, \citenamefont {Jeong}, \citenamefont {Ahn},
  \citenamefont {Chang}, \citenamefont {Vlassopoulos}, \citenamefont
  {Halverson}, \citenamefont {Kremer}, \citenamefont {Ge}, \citenamefont
  {Rubinstein}, \citenamefont {Grest}, \citenamefont {Srinin},\ and\
  \citenamefont {Grosberg}}]{parisi2021Nonlinear}%
  \BibitemOpen
  \bibfield  {author} {\bibinfo {author} {\bibfnamefont {D.}~\bibnamefont
  {Parisi}}, \bibinfo {author} {\bibfnamefont {S.}~\bibnamefont {Costanzo}},
  \bibinfo {author} {\bibfnamefont {Y.}~\bibnamefont {Jeong}}, \bibinfo
  {author} {\bibfnamefont {J.}~\bibnamefont {Ahn}}, \bibinfo {author}
  {\bibfnamefont {T.}~\bibnamefont {Chang}}, \bibinfo {author} {\bibfnamefont
  {D.}~\bibnamefont {Vlassopoulos}}, \bibinfo {author} {\bibfnamefont {J.~D.}\
  \bibnamefont {Halverson}}, \bibinfo {author} {\bibfnamefont {K.}~\bibnamefont
  {Kremer}}, \bibinfo {author} {\bibfnamefont {T.}~\bibnamefont {Ge}}, \bibinfo
  {author} {\bibfnamefont {M.}~\bibnamefont {Rubinstein}}, \bibinfo {author}
  {\bibfnamefont {G.~S.}\ \bibnamefont {Grest}}, \bibinfo {author}
  {\bibfnamefont {W.}~\bibnamefont {Srinin}},\ and\ \bibinfo {author}
  {\bibfnamefont {A.~Y.}\ \bibnamefont {Grosberg}},\ }\bibfield  {title}
  {\bibinfo {title} {Nonlinear {{Shear Rheology}} of {{Entangled Polymer
  Rings}}},\ }\href {https://doi.org/10.1021/acs.macromol.0c02839} {\bibfield
  {journal} {\bibinfo  {journal} {Macromolecules}\ }\textbf {\bibinfo {volume}
  {54}},\ \bibinfo {pages} {2811} (\bibinfo {year} {2021})}\BibitemShut
  {NoStop}%
\bibitem [{\citenamefont {Tu}\ \emph {et~al.}(2023)\citenamefont {Tu},
  \citenamefont {Davydovich}, \citenamefont {Mei}, \citenamefont {Singh},
  \citenamefont {Grest}, \citenamefont {Schweizer}, \citenamefont {O'Connor},\
  and\ \citenamefont {Schroeder}}]{tu2023Unexpected}%
  \BibitemOpen
  \bibfield  {author} {\bibinfo {author} {\bibfnamefont {M.~Q.}\ \bibnamefont
  {Tu}}, \bibinfo {author} {\bibfnamefont {O.}~\bibnamefont {Davydovich}},
  \bibinfo {author} {\bibfnamefont {B.}~\bibnamefont {Mei}}, \bibinfo {author}
  {\bibfnamefont {P.~K.}\ \bibnamefont {Singh}}, \bibinfo {author}
  {\bibfnamefont {G.~S.}\ \bibnamefont {Grest}}, \bibinfo {author}
  {\bibfnamefont {K.~S.}\ \bibnamefont {Schweizer}}, \bibinfo {author}
  {\bibfnamefont {T.~C.}\ \bibnamefont {O'Connor}},\ and\ \bibinfo {author}
  {\bibfnamefont {C.~M.}\ \bibnamefont {Schroeder}},\ }\bibfield  {title}
  {\bibinfo {title} {Unexpected {{Slow Relaxation Dynamics}} in {{Pure Ring
  Polymers Arise}} from {{Intermolecular Interactions}}},\ }\bibfield
  {journal} {\bibinfo  {journal} {ACS Polym. Au}\ }\href
  {https://doi.org/10.1021/acspolymersau.2c00069}
  {10.1021/acspolymersau.2c00069} (\bibinfo {year} {2023})\BibitemShut
  {NoStop}%
\bibitem [{\citenamefont {Faller}\ \emph {et~al.}(1999)\citenamefont {Faller},
  \citenamefont {Kolb},\ and\ \citenamefont
  {{M{\"u}ller-Plathe}}}]{faller1999Local}%
  \BibitemOpen
  \bibfield  {author} {\bibinfo {author} {\bibfnamefont {R.}~\bibnamefont
  {Faller}}, \bibinfo {author} {\bibfnamefont {A.}~\bibnamefont {Kolb}},\ and\
  \bibinfo {author} {\bibfnamefont {F.}~\bibnamefont {{M{\"u}ller-Plathe}}},\
  }\bibfield  {title} {\bibinfo {title} {Local chain ordering in amorphous
  polymer melts: Influence of chain stiffness},\ }\href
  {https://doi.org/10.1039/A809796H} {\bibfield  {journal} {\bibinfo  {journal}
  {Phys. Chem. Chem. Phys.}\ }\textbf {\bibinfo {volume} {1}},\ \bibinfo
  {pages} {2071} (\bibinfo {year} {1999})}\BibitemShut {NoStop}%
\bibitem [{\citenamefont {Sukumaran}\ \emph {et~al.}(2005)\citenamefont
  {Sukumaran}, \citenamefont {Grest}, \citenamefont {Kremer},\ and\
  \citenamefont {Everaers}}]{sukumaran2005Identifying}%
  \BibitemOpen
  \bibfield  {author} {\bibinfo {author} {\bibfnamefont {S.~K.}\ \bibnamefont
  {Sukumaran}}, \bibinfo {author} {\bibfnamefont {G.~S.}\ \bibnamefont
  {Grest}}, \bibinfo {author} {\bibfnamefont {K.}~\bibnamefont {Kremer}},\ and\
  \bibinfo {author} {\bibfnamefont {R.}~\bibnamefont {Everaers}},\ }\bibfield
  {title} {\bibinfo {title} {Identifying the primitive path mesh in entangled
  polymer liquids},\ }\href {https://doi.org/10.1002/polb.20384} {\bibfield
  {journal} {\bibinfo  {journal} {J. Polym. Sci. B Polym. Phys.}\ }\textbf
  {\bibinfo {volume} {43}},\ \bibinfo {pages} {917} (\bibinfo {year}
  {2005})}\BibitemShut {NoStop}%
\bibitem [{\citenamefont {Hagita}\ and\ \citenamefont
  {Murashima}(2021)}]{hagita2021Effect}%
  \BibitemOpen
  \bibfield  {author} {\bibinfo {author} {\bibfnamefont {K.}~\bibnamefont
  {Hagita}}\ and\ \bibinfo {author} {\bibfnamefont {T.}~\bibnamefont
  {Murashima}},\ }\bibfield  {title} {\bibinfo {title} {Effect of
  chain-penetration on ring shape for mixtures of rings and linear polymers},\
  }\href {https://doi.org/10.1016/j.polymer.2021.123493} {\bibfield  {journal}
  {\bibinfo  {journal} {Polymer}\ }\textbf {\bibinfo {volume} {218}},\ \bibinfo
  {pages} {123493} (\bibinfo {year} {2021})}\BibitemShut {NoStop}%
\bibitem [{\citenamefont {Everaers}\ \emph {et~al.}(2004)\citenamefont
  {Everaers}, \citenamefont {Sukumaran}, \citenamefont {Grest}, \citenamefont
  {Svaneborg}, \citenamefont {Sivasubramanian},\ and\ \citenamefont
  {Kremer}}]{everaers2004Rheology}%
  \BibitemOpen
  \bibfield  {author} {\bibinfo {author} {\bibfnamefont {R.}~\bibnamefont
  {Everaers}}, \bibinfo {author} {\bibfnamefont {S.~K.}\ \bibnamefont
  {Sukumaran}}, \bibinfo {author} {\bibfnamefont {G.~S.}\ \bibnamefont
  {Grest}}, \bibinfo {author} {\bibfnamefont {C.}~\bibnamefont {Svaneborg}},
  \bibinfo {author} {\bibfnamefont {A.}~\bibnamefont {Sivasubramanian}},\ and\
  \bibinfo {author} {\bibfnamefont {K.}~\bibnamefont {Kremer}},\ }\bibfield
  {title} {\bibinfo {title} {Rheology and {{Microscopic Topology}} of
  {{Entangled Polymeric Liquids}}},\ }\href
  {https://doi.org/10.1126/science.1091215} {\bibfield  {journal} {\bibinfo
  {journal} {Science}\ }\textbf {\bibinfo {volume} {303}},\ \bibinfo {pages}
  {823} (\bibinfo {year} {2004})}\BibitemShut {NoStop}%
\bibitem [{\citenamefont {Aronovitz}\ and\ \citenamefont
  {Nelson}(1986)}]{aronovitz1986Universal}%
  \BibitemOpen
  \bibfield  {author} {\bibinfo {author} {\bibfnamefont {J.}~\bibnamefont
  {Aronovitz}}\ and\ \bibinfo {author} {\bibfnamefont {D.}~\bibnamefont
  {Nelson}},\ }\bibfield  {title} {\bibinfo {title} {Universal features of
  polymer shapes},\ }\href {https://doi.org/10.1051/jphys:019860047090144500}
  {\bibfield  {journal} {\bibinfo  {journal} {J. Phys. France}\ }\textbf
  {\bibinfo {volume} {47}},\ \bibinfo {pages} {1445} (\bibinfo {year}
  {1986})}\BibitemShut {NoStop}%
\bibitem [{\citenamefont {Rudnick}\ and\ \citenamefont
  {Gaspari}(1986)}]{rudnick1986Aspherity}%
  \BibitemOpen
  \bibfield  {author} {\bibinfo {author} {\bibfnamefont {J.}~\bibnamefont
  {Rudnick}}\ and\ \bibinfo {author} {\bibfnamefont {G.}~\bibnamefont
  {Gaspari}},\ }\bibfield  {title} {\bibinfo {title} {The aspherity of random
  walks},\ }\href {https://doi.org/10.1088/0305-4470/19/4/004} {\bibfield
  {journal} {\bibinfo  {journal} {J. Phys. A: Math. Gen.}\ }\textbf {\bibinfo
  {volume} {19}},\ \bibinfo {pages} {L191} (\bibinfo {year}
  {1986})}\BibitemShut {NoStop}%
\bibitem [{\citenamefont {Gaspari}\ \emph {et~al.}(1987)\citenamefont
  {Gaspari}, \citenamefont {Rudnick},\ and\ \citenamefont
  {Beldjenna}}]{gaspari1987Shapes}%
  \BibitemOpen
  \bibfield  {author} {\bibinfo {author} {\bibfnamefont {G.}~\bibnamefont
  {Gaspari}}, \bibinfo {author} {\bibfnamefont {J.}~\bibnamefont {Rudnick}},\
  and\ \bibinfo {author} {\bibfnamefont {A.}~\bibnamefont {Beldjenna}},\
  }\bibfield  {title} {\bibinfo {title} {The shapes of open and closed random
  walks: A 1/d expansion},\ }\href
  {https://doi.org/10.1088/0305-4470/20/11/041} {\bibfield  {journal} {\bibinfo
   {journal} {J. Phys. A: Math. Gen.}\ }\textbf {\bibinfo {volume} {20}},\
  \bibinfo {pages} {3393} (\bibinfo {year} {1987})}\BibitemShut {NoStop}%
\bibitem [{\citenamefont {Jagodzinski}\ \emph {et~al.}(1992)\citenamefont
  {Jagodzinski}, \citenamefont {Eisenriegler},\ and\ \citenamefont
  {Kremer}}]{jagodzinski1992Universal}%
  \BibitemOpen
  \bibfield  {author} {\bibinfo {author} {\bibfnamefont {O.}~\bibnamefont
  {Jagodzinski}}, \bibinfo {author} {\bibfnamefont {E.}~\bibnamefont
  {Eisenriegler}},\ and\ \bibinfo {author} {\bibfnamefont {K.}~\bibnamefont
  {Kremer}},\ }\bibfield  {title} {\bibinfo {title} {Universal shape properties
  of open and closed polymer chains: Renormalization group analysis and {{Monte
  Carlo}} experiments},\ }\href {https://doi.org/10.1051/jp1:1992279}
  {\bibfield  {journal} {\bibinfo  {journal} {J. Phys. I France}\ }\textbf
  {\bibinfo {volume} {2}},\ \bibinfo {pages} {2243} (\bibinfo {year}
  {1992})}\BibitemShut {NoStop}%
\bibitem [{\citenamefont {Bishop}\ and\ \citenamefont
  {Michels}(1985)}]{bishop1985Shape}%
  \BibitemOpen
  \bibfield  {author} {\bibinfo {author} {\bibfnamefont {M.}~\bibnamefont
  {Bishop}}\ and\ \bibinfo {author} {\bibfnamefont {J.~P.~J.}\ \bibnamefont
  {Michels}},\ }\bibfield  {title} {\bibinfo {title} {The shape of ring
  polymers},\ }\href {https://doi.org/10.1063/1.448949} {\bibfield  {journal}
  {\bibinfo  {journal} {J. Chem. Phys.}\ }\textbf {\bibinfo {volume} {82}},\
  \bibinfo {pages} {1059} (\bibinfo {year} {1985})}\BibitemShut {NoStop}%
\bibitem [{\citenamefont {Bishop}\ and\ \citenamefont
  {Saltiel}(1988)}]{bishop1988Polymer}%
  \BibitemOpen
  \bibfield  {author} {\bibinfo {author} {\bibfnamefont {M.}~\bibnamefont
  {Bishop}}\ and\ \bibinfo {author} {\bibfnamefont {C.~J.}\ \bibnamefont
  {Saltiel}},\ }\bibfield  {title} {\bibinfo {title} {Polymer shapes in two,
  four, and five dimensions},\ }\href {https://doi.org/10.1063/1.453847}
  {\bibfield  {journal} {\bibinfo  {journal} {J. Chem. Phys.}\ }\textbf
  {\bibinfo {volume} {88}},\ \bibinfo {pages} {3976} (\bibinfo {year}
  {1988})}\BibitemShut {NoStop}%
\bibitem [{\citenamefont {Zifferer}\ and\ \citenamefont
  {Preusser}(2001)}]{zifferer2001Monte}%
  \BibitemOpen
  \bibfield  {author} {\bibinfo {author} {\bibfnamefont {G.}~\bibnamefont
  {Zifferer}}\ and\ \bibinfo {author} {\bibfnamefont {W.}~\bibnamefont
  {Preusser}},\ }\bibfield  {title} {\bibinfo {title} {Monte {{Carlo Simulation
  Studies}} of the {{Size}} and {{Shape}} of {{Ring Polymers}}},\ }\href
  {https://doi.org/10.1002/1521-3919(20010601)10:5<397::AID-MATS397>3.0.CO;2-X}
  {\bibfield  {journal} {\bibinfo  {journal} {Macromol. Theory Simul.}\
  }\textbf {\bibinfo {volume} {10}},\ \bibinfo {pages} {397} (\bibinfo {year}
  {2001})}\BibitemShut {NoStop}%
\bibitem [{\citenamefont {Alim}\ and\ \citenamefont
  {Frey}(2007)}]{alim2007Shapes}%
  \BibitemOpen
  \bibfield  {author} {\bibinfo {author} {\bibfnamefont {K.}~\bibnamefont
  {Alim}}\ and\ \bibinfo {author} {\bibfnamefont {E.}~\bibnamefont {Frey}},\
  }\bibfield  {title} {\bibinfo {title} {Shapes of {{Semiflexible Polymer
  Rings}}},\ }\href {https://doi.org/10.1103/PhysRevLett.99.198102} {\bibfield
  {journal} {\bibinfo  {journal} {Phys. Rev. Lett.}\ }\textbf {\bibinfo
  {volume} {99}},\ \bibinfo {pages} {198102} (\bibinfo {year}
  {2007})}\BibitemShut {NoStop}%
\bibitem [{\citenamefont {Rawdon}\ \emph {et~al.}(2008)\citenamefont {Rawdon},
  \citenamefont {Kern}, \citenamefont {Piatek}, \citenamefont {Plunkett},
  \citenamefont {Stasiak},\ and\ \citenamefont {Millett}}]{rawdon2008Effect}%
  \BibitemOpen
  \bibfield  {author} {\bibinfo {author} {\bibfnamefont {E.~J.}\ \bibnamefont
  {Rawdon}}, \bibinfo {author} {\bibfnamefont {J.~C.}\ \bibnamefont {Kern}},
  \bibinfo {author} {\bibfnamefont {M.}~\bibnamefont {Piatek}}, \bibinfo
  {author} {\bibfnamefont {P.}~\bibnamefont {Plunkett}}, \bibinfo {author}
  {\bibfnamefont {A.}~\bibnamefont {Stasiak}},\ and\ \bibinfo {author}
  {\bibfnamefont {K.~C.}\ \bibnamefont {Millett}},\ }\bibfield  {title}
  {\bibinfo {title} {Effect of {{Knotting}} on the {{Shape}} of {{Polymers}}},\
  }\href {https://doi.org/10.1021/ma801389c} {\bibfield  {journal} {\bibinfo
  {journal} {Macromolecules}\ }\textbf {\bibinfo {volume} {41}},\ \bibinfo
  {pages} {8281} (\bibinfo {year} {2008})}\BibitemShut {NoStop}%
\bibitem [{\citenamefont {Reigh}\ and\ \citenamefont
  {Yoon}(2013)}]{reigh2013Concentration}%
  \BibitemOpen
  \bibfield  {author} {\bibinfo {author} {\bibfnamefont {S.~Y.}\ \bibnamefont
  {Reigh}}\ and\ \bibinfo {author} {\bibfnamefont {D.~Y.}\ \bibnamefont
  {Yoon}},\ }\bibfield  {title} {\bibinfo {title} {Concentration {{Dependence}}
  of {{Ring Polymer Conformations}} from {{Monte Carlo Simulations}}},\ }\href
  {https://doi.org/10.1021/mz300587v} {\bibfield  {journal} {\bibinfo
  {journal} {ACS Macro Lett.}\ }\textbf {\bibinfo {volume} {2}},\ \bibinfo
  {pages} {296} (\bibinfo {year} {2013})}\BibitemShut {NoStop}%
\bibitem [{\citenamefont {Cai}\ \emph {et~al.}(2022)\citenamefont {Cai},
  \citenamefont {Liang}, \citenamefont {Liu},\ and\ \citenamefont
  {Zhang}}]{cai2022Conformation}%
  \BibitemOpen
  \bibfield  {author} {\bibinfo {author} {\bibfnamefont {X.}~\bibnamefont
  {Cai}}, \bibinfo {author} {\bibfnamefont {C.}~\bibnamefont {Liang}}, \bibinfo
  {author} {\bibfnamefont {H.}~\bibnamefont {Liu}},\ and\ \bibinfo {author}
  {\bibfnamefont {G.}~\bibnamefont {Zhang}},\ }\bibfield  {title} {\bibinfo
  {title} {Conformation and structure of ring polymers in semidilute solutions:
  {{A}} molecular dynamics simulation study},\ }\href
  {https://doi.org/10.1016/j.polymer.2022.124953} {\bibfield  {journal}
  {\bibinfo  {journal} {Polymer}\ }\textbf {\bibinfo {volume} {253}},\ \bibinfo
  {pages} {124953} (\bibinfo {year} {2022})}\BibitemShut {NoStop}%
\bibitem [{\citenamefont {Michieletto}\ \emph {et~al.}(2015)\citenamefont
  {Michieletto}, \citenamefont {Marenduzzo},\ and\ \citenamefont
  {Orlandini}}]{michieletto2015Kinetoplast}%
  \BibitemOpen
  \bibfield  {author} {\bibinfo {author} {\bibfnamefont {D.}~\bibnamefont
  {Michieletto}}, \bibinfo {author} {\bibfnamefont {D.}~\bibnamefont
  {Marenduzzo}},\ and\ \bibinfo {author} {\bibfnamefont {E.}~\bibnamefont
  {Orlandini}},\ }\bibfield  {title} {\bibinfo {title} {Is the kinetoplast
  {{DNA}} a percolating network of linked rings at its critical point?},\
  }\href {https://doi.org/10.1088/1478-3975/12/3/036001} {\bibfield  {journal}
  {\bibinfo  {journal} {Phys. Biol.}\ }\textbf {\bibinfo {volume} {12}},\
  \bibinfo {pages} {036001} (\bibinfo {year} {2015})}\BibitemShut {NoStop}%
\bibitem [{\citenamefont {Yamamoto}\ and\ \citenamefont
  {Onuki}(1998)}]{yamamoto1998Dynamics}%
  \BibitemOpen
  \bibfield  {author} {\bibinfo {author} {\bibfnamefont {R.}~\bibnamefont
  {Yamamoto}}\ and\ \bibinfo {author} {\bibfnamefont {A.}~\bibnamefont
  {Onuki}},\ }\bibfield  {title} {\bibinfo {title} {Dynamics of highly
  supercooled liquids: {{Heterogeneity}}, rheology, and diffusion},\ }\href
  {https://doi.org/10.1103/PhysRevE.58.3515} {\bibfield  {journal} {\bibinfo
  {journal} {Phys. Rev. E}\ }\textbf {\bibinfo {volume} {58}},\ \bibinfo
  {pages} {3515} (\bibinfo {year} {1998})}\BibitemShut {NoStop}%
\bibitem [{\citenamefont {Shiba}\ \emph {et~al.}(2012)\citenamefont {Shiba},
  \citenamefont {Kawasaki},\ and\ \citenamefont
  {Onuki}}]{shiba2012Relationship}%
  \BibitemOpen
  \bibfield  {author} {\bibinfo {author} {\bibfnamefont {H.}~\bibnamefont
  {Shiba}}, \bibinfo {author} {\bibfnamefont {T.}~\bibnamefont {Kawasaki}},\
  and\ \bibinfo {author} {\bibfnamefont {A.}~\bibnamefont {Onuki}},\ }\bibfield
   {title} {\bibinfo {title} {Relationship between bond-breakage correlations
  and four-point correlations in heterogeneous glassy dynamics:
  {{Configuration}} changes and vibration modes},\ }\href
  {https://doi.org/10.1103/PhysRevE.86.041504} {\bibfield  {journal} {\bibinfo
  {journal} {Phys. Rev. E}\ }\textbf {\bibinfo {volume} {86}},\ \bibinfo
  {pages} {041504} (\bibinfo {year} {2012})}\BibitemShut {NoStop}%
\bibitem [{\citenamefont {Kawasaki}\ and\ \citenamefont
  {Onuki}(2013)}]{kawasaki2013Slow}%
  \BibitemOpen
  \bibfield  {author} {\bibinfo {author} {\bibfnamefont {T.}~\bibnamefont
  {Kawasaki}}\ and\ \bibinfo {author} {\bibfnamefont {A.}~\bibnamefont
  {Onuki}},\ }\bibfield  {title} {\bibinfo {title} {Slow relaxations and
  stringlike jump motions in fragile glass-forming liquids: {{Breakdown}} of
  the {{Stokes-Einstein}} relation},\ }\href
  {https://doi.org/10.1103/PhysRevE.87.012312} {\bibfield  {journal} {\bibinfo
  {journal} {Phys. Rev. E}\ }\textbf {\bibinfo {volume} {87}},\ \bibinfo
  {pages} {012312} (\bibinfo {year} {2013})}\BibitemShut {NoStop}%
\bibitem [{\citenamefont {Shiba}\ \emph {et~al.}(2016)\citenamefont {Shiba},
  \citenamefont {Yamada}, \citenamefont {Kawasaki},\ and\ \citenamefont
  {Kim}}]{shiba2016Unveiling}%
  \BibitemOpen
  \bibfield  {author} {\bibinfo {author} {\bibfnamefont {H.}~\bibnamefont
  {Shiba}}, \bibinfo {author} {\bibfnamefont {Y.}~\bibnamefont {Yamada}},
  \bibinfo {author} {\bibfnamefont {T.}~\bibnamefont {Kawasaki}},\ and\
  \bibinfo {author} {\bibfnamefont {K.}~\bibnamefont {Kim}},\ }\bibfield
  {title} {\bibinfo {title} {Unveiling {{Dimensionality Dependence}} of
  {{Glassy Dynamics}}: {{2D Infinite Fluctuation Eclipses Inherent Structural
  Relaxation}}},\ }\href {https://doi.org/10.1103/PhysRevLett.117.245701}
  {\bibfield  {journal} {\bibinfo  {journal} {Phys. Rev. Lett.}\ }\textbf
  {\bibinfo {volume} {117}},\ \bibinfo {pages} {245701} (\bibinfo {year}
  {2016})}\BibitemShut {NoStop}%
\bibitem [{\citenamefont {Dell}\ and\ \citenamefont
  {Schweizer}(2018)}]{dell2018Intermolecular}%
  \BibitemOpen
  \bibfield  {author} {\bibinfo {author} {\bibfnamefont {Z.~E.}\ \bibnamefont
  {Dell}}\ and\ \bibinfo {author} {\bibfnamefont {K.~S.}\ \bibnamefont
  {Schweizer}},\ }\bibfield  {title} {\bibinfo {title} {Intermolecular
  structural correlations in model globular and unconcatenated ring polymer
  liquids},\ }\href {https://doi.org/10.1039/C8SM01722K} {\bibfield  {journal}
  {\bibinfo  {journal} {Soft Matter}\ }\textbf {\bibinfo {volume} {14}},\
  \bibinfo {pages} {9132} (\bibinfo {year} {2018})}\BibitemShut {NoStop}%
\bibitem [{\citenamefont {Mei}\ \emph {et~al.}(2020)\citenamefont {Mei},
  \citenamefont {Dell},\ and\ \citenamefont {Schweizer}}]{mei2020Microscopic}%
  \BibitemOpen
  \bibfield  {author} {\bibinfo {author} {\bibfnamefont {B.}~\bibnamefont
  {Mei}}, \bibinfo {author} {\bibfnamefont {Z.~E.}\ \bibnamefont {Dell}},\ and\
  \bibinfo {author} {\bibfnamefont {K.~S.}\ \bibnamefont {Schweizer}},\
  }\bibfield  {title} {\bibinfo {title} {Microscopic {{Theory}} of {{Long-Time
  Center-of-Mass Self-Diffusion}} and {{Anomalous Transport}} in {{Ring Polymer
  Liquids}}},\ }\href {https://doi.org/10.1021/acs.macromol.0c01737} {\bibfield
   {journal} {\bibinfo  {journal} {Macromolecules}\ }\textbf {\bibinfo {volume}
  {53}},\ \bibinfo {pages} {10431} (\bibinfo {year} {2020})}\BibitemShut
  {NoStop}%
\bibitem [{\citenamefont {Mei}\ \emph {et~al.}(2021)\citenamefont {Mei},
  \citenamefont {Dell},\ and\ \citenamefont {Schweizer}}]{mei2021Theorya}%
  \BibitemOpen
  \bibfield  {author} {\bibinfo {author} {\bibfnamefont {B.}~\bibnamefont
  {Mei}}, \bibinfo {author} {\bibfnamefont {Z.~E.}\ \bibnamefont {Dell}},\ and\
  \bibinfo {author} {\bibfnamefont {K.~S.}\ \bibnamefont {Schweizer}},\
  }\bibfield  {title} {\bibinfo {title} {Theory of {{Transient Localization}},
  {{Activated Dynamics}}, and a {{Macromolecular Glass Transition}} in {{Ring
  Polymer Liquids}}},\ }\href {https://doi.org/10.1021/acsmacrolett.1c00530}
  {\bibfield  {journal} {\bibinfo  {journal} {ACS Macro Lett.}\ }\textbf
  {\bibinfo {volume} {10}},\ \bibinfo {pages} {1229} (\bibinfo {year}
  {2021})}\BibitemShut {NoStop}%
\bibitem [{\citenamefont {Jeong}\ and\ \citenamefont
  {Douglas}(2017)}]{jeong2017Relationa}%
  \BibitemOpen
  \bibfield  {author} {\bibinfo {author} {\bibfnamefont {C.}~\bibnamefont
  {Jeong}}\ and\ \bibinfo {author} {\bibfnamefont {J.~F.}\ \bibnamefont
  {Douglas}},\ }\bibfield  {title} {\bibinfo {title} {Relation between
  {{Polymer Conformational Structure}} and {{Dynamics}} in {{Linear}} and
  {{Ring Polyethylene Blends}}},\ }\href
  {https://doi.org/10.1002/mats.201700045} {\bibfield  {journal} {\bibinfo
  {journal} {Macromol. Theory Simul.}\ }\textbf {\bibinfo {volume} {26}},\
  \bibinfo {pages} {1700045} (\bibinfo {year} {2017})}\BibitemShut {NoStop}%
\bibitem [{\citenamefont {Borger}\ \emph {et~al.}(2020)\citenamefont {Borger},
  \citenamefont {Wang}, \citenamefont {O'Connor}, \citenamefont {Ge},
  \citenamefont {Grest}, \citenamefont {Jensen}, \citenamefont {Ahn},
  \citenamefont {Chang}, \citenamefont {Hassager}, \citenamefont {Mortensen},
  \citenamefont {Vlassopoulos},\ and\ \citenamefont
  {Huang}}]{borger2020Threading}%
  \BibitemOpen
  \bibfield  {author} {\bibinfo {author} {\bibfnamefont {A.}~\bibnamefont
  {Borger}}, \bibinfo {author} {\bibfnamefont {W.}~\bibnamefont {Wang}},
  \bibinfo {author} {\bibfnamefont {T.~C.}\ \bibnamefont {O'Connor}}, \bibinfo
  {author} {\bibfnamefont {T.}~\bibnamefont {Ge}}, \bibinfo {author}
  {\bibfnamefont {G.~S.}\ \bibnamefont {Grest}}, \bibinfo {author}
  {\bibfnamefont {G.~V.}\ \bibnamefont {Jensen}}, \bibinfo {author}
  {\bibfnamefont {J.}~\bibnamefont {Ahn}}, \bibinfo {author} {\bibfnamefont
  {T.}~\bibnamefont {Chang}}, \bibinfo {author} {\bibfnamefont
  {O.}~\bibnamefont {Hassager}}, \bibinfo {author} {\bibfnamefont
  {K.}~\bibnamefont {Mortensen}}, \bibinfo {author} {\bibfnamefont
  {D.}~\bibnamefont {Vlassopoulos}},\ and\ \bibinfo {author} {\bibfnamefont
  {Q.}~\bibnamefont {Huang}},\ }\bibfield  {title} {\bibinfo {title}
  {Threading\textendash{{Unthreading Transition}} of {{Linear-Ring Polymer
  Blends}} in {{Extensional Flow}}},\ }\href
  {https://doi.org/10.1021/acsmacrolett.0c00607} {\bibfield  {journal}
  {\bibinfo  {journal} {ACS Macro Lett.}\ }\textbf {\bibinfo {volume} {9}},\
  \bibinfo {pages} {1452} (\bibinfo {year} {2020})}\BibitemShut {NoStop}%
\bibitem [{\citenamefont {O'Connor}\ \emph {et~al.}(2022)\citenamefont
  {O'Connor}, \citenamefont {Ge},\ and\ \citenamefont
  {Grest}}]{oconnor2022Composite}%
  \BibitemOpen
  \bibfield  {author} {\bibinfo {author} {\bibfnamefont {T.~C.}\ \bibnamefont
  {O'Connor}}, \bibinfo {author} {\bibfnamefont {T.}~\bibnamefont {Ge}},\ and\
  \bibinfo {author} {\bibfnamefont {G.~S.}\ \bibnamefont {Grest}},\ }\bibfield
  {title} {\bibinfo {title} {Composite entanglement topology and extensional
  rheology of symmetric ring-linear polymer blends},\ }\href
  {https://doi.org/10.1122/8.0000319} {\bibfield  {journal} {\bibinfo
  {journal} {J. Rheol.}\ }\textbf {\bibinfo {volume} {66}},\ \bibinfo {pages}
  {49} (\bibinfo {year} {2022})}\BibitemShut {NoStop}%
\bibitem [{\citenamefont {Grest}\ \emph {et~al.}(2023)\citenamefont {Grest},
  \citenamefont {Ge}, \citenamefont {Plimpton}, \citenamefont {Rubinstein},\
  and\ \citenamefont {O'Connor}}]{grest2023Entropic}%
  \BibitemOpen
  \bibfield  {author} {\bibinfo {author} {\bibfnamefont {G.~S.}\ \bibnamefont
  {Grest}}, \bibinfo {author} {\bibfnamefont {T.}~\bibnamefont {Ge}}, \bibinfo
  {author} {\bibfnamefont {S.~J.}\ \bibnamefont {Plimpton}}, \bibinfo {author}
  {\bibfnamefont {M.}~\bibnamefont {Rubinstein}},\ and\ \bibinfo {author}
  {\bibfnamefont {T.~C.}\ \bibnamefont {O'Connor}},\ }\bibfield  {title}
  {\bibinfo {title} {Entropic {{Mixing}} of {{Ring}}/{{Linear Polymer
  Blends}}},\ }\href {https://doi.org/10.1021/acspolymersau.2c00050} {\bibfield
   {journal} {\bibinfo  {journal} {ACS Polym. Au}\ }\textbf {\bibinfo {volume}
  {3}},\ \bibinfo {pages} {209} (\bibinfo {year} {2023})}\BibitemShut {NoStop}%
\end{thebibliography}

\end{document}